\UseRawInputEncoding
\documentclass[sn-mathphys]{sn-jnl}

\usepackage[utf8]{inputenc}
\usepackage{longtable}
\usepackage{graphicx}%
\usepackage{multirow}%
\usepackage{amsmath,amssymb,amsfonts}%
\usepackage{amsthm}%
\usepackage{mathrsfs}%
\usepackage[title]{appendix}%
\usepackage{xcolor}%
\usepackage{textcomp}%
\usepackage{natbib}%
\usepackage{manyfoot}%
\usepackage{booktabs}%
\usepackage{algorithm}%
\usepackage{algorithmicx}%
\usepackage{algpseudocode}%
\usepackage{listings}%
\usepackage{array}%
\newcolumntype{P}[1]{>{\raggedright\arraybackslash}p{#1}}
\usepackage{float}
\usepackage{afterpage,float}

\DeclareUnicodeCharacter{202F}{\,}
\begin{document}

\title{The QTF-Backbone: Proposal for a Nationwide Optical Fibre Backbone in Germany for Quantum Technology and Time and Frequency Metrology}
\author[1]{\fnm{Tara Cubel} \sur{Liebisch}}
\author[2]{\fnm{Peter} \sur{Kaufmann}}
\author[1]{\fnm{Harald} \sur{Schnatz}}
\author[3]{\fnm{Susanne} \sur{Naegele-Jackson}}
\author[1]{\fnm{Jochen} \sur{Kronjäger}}
\author[4]{\fnm{Klaus} \sur{Blaum}}
\author[1]{\fnm{Stefan} \sur{Kück}}
\author[5]{\fnm{Dieter} \sur{Meschede}}
\author[6]{\fnm{Stephan} \sur{Schiller}}
\author[7]{\fnm{Laura} \sur{Agazzi}}
% ...

\affil[1]{Physikalisch-Technische Bundesanstalt}
\affil[2]{German National Research and Education Network, DFN}
\affil[3]{Friedrich-Alexander-Universität Erlangen-Nürnberg}
\affil[4]{Max Planck Institute for Nuclear Physics}
\affil[5]{University of Bonn}
\affil[6]{Heinrich Heine University Düsseldorf}
\affil[7]{German Aerospace Center, DLR}
% ...


\date{May 22nd 2026}

\maketitle
\newpage

\section*{Abstract}

The recent breakthroughs in the distribution of quantum information and high-precision time and frequency (T\&F) signals over long-haul optical fibre networks have transformative potential for physically secure communications, resilience of timing infrastructure (such as that supporting Global Navigation Satellite Systems (GNSS)) and fundamental physics. To date, these capabilities remain confined to isolated testbeds, with quantum and T\&F signals accessible, for example in Germany, to only a few institutions.

In this white paper we propose the QTF-Backbone: a dedicated national fibre-optic infrastructure in Germany for the networked distribution of \textbf{Q}uantum and \textbf{T\&F} signals using dark fibres and specialised hardware. The QTF-Backbone is planned as a four-phase deployment over ten years to ensure scalable, sustainable access for research institutions and industry. The concept builds on successful demonstrations of time and frequency distribution at high Technology Readiness Levels (TRLs) across Europe, including PTB--MPQ links in Germany, REFIMEVE in France, and the Italian LIFT network. The QTF-Backbone will enable transformative Research and Development (R\&D), support a nationwide QTF ecosystem, and ensure the transition from innovation to deployment. As a national and European hub, it will position Germany and Europe at the forefront of quantum networking, as well as T\&F transfer.

\section*{Keywords}

fibre-optic network, R\&D infrastructure, quantum links, time-frequency transfer, metrology, distributed quantum computing, precision measurements, quantum sensing, quantum computing, quantum communication, quantum technologies, navigation, geosciences, height reference systems, resilience.

\newpage

\section{Introduction}

Two key technologies that have emerged in optical fibre networks are quantum communication and the distribution of high-precision time and frequency (T\&F) signals. These advancements open the door to transformative societal impact, including Research and Development (R\&D) for physically secure message transmission,
and resilience of timing infrastructure~\cite{Clivati2022}(and references [5,12,32,38–40] therein),~\cite{Goward2023}, such as that supporting Global Navigation Satellite Systems (GNSS). In the opposite way, GNSS-based methods are widely used for the generation and the dissemination of Coordinated Universal Time (UTC)~\cite{Bauch2020}. However, to unlock the full potential of these applications it is crucial to move beyond sparse and isolated testbed fibre links toward an interconnected, nationwide fibre optic network infrastructure, making these capabilities accessible to a large range of users. For example, specialised ground stations enabling space-to-ground Quantum Key Distribution (QKD) networks, trusted nodes, quantum computers and first-generation quantum repeaters are accessible to limited numbers of research institutions and companies in Germany. Similarly, the highest-performance T\&F signals are currently available in Germany only at the National Metrology Institute (NMI), the Physikalisch-Technische Bundesanstalt (PTB), and at a few laboratories connected via fibre links that lack guaranteed long-term funding. These technologies are now ready for deployment in a national large-scale research infrastructure (RI) that links and advances the capabilities of stakeholders across academia, government, and industry. 

We propose a national fibre optic network, a ``QTF-Backbone'' to connect and provide research facilities across Germany with a dedicated infrastructure for experimental quantum links (\textbf{Q}) alongside time and frequency (\textbf{T\&F}) distribution. The QTF-Backbone can be installed and operation can be established in four phases over 10 years to provide increasing and sustained user access to these two groundbreaking technologies and to create new R\&D opportunities that existing networks in Germany cannot provide. For example, the common Research \& Education (R\&E)-X-WiN-network from DFN, Germany's National Research and Education Network (NREN), or special purpose networks like the core transport network of the federal government (KTN) from the Federal Ministry of the Interior and Community (BMI) or the SaSER network provided by  the Deutsche Telekom AG, are currently unable to offer the combination of these technologies, as they present a particular technical challenge for network transmission that can only be met by deploying dedicated hardware along dark fibres at strategic locations.

The QTF-Backbone will enable R\&D in the field applications -- outside of controlled laboratory environments -- for quantum communication, quantum computing, quantum sensing and other T\&F-critical applications across academia and industry. The synergistic benefits of high-performance quantum links and T\&F signals can be explored and rapidly advanced. The QTF-Backbone will also enable unprecedented research on sensor networks for investigating fundamental physics, environmental monitoring, astronomical and geodetic observations, and infrastructural applications. This RI will act as a national collaboration platform across disciplines and institutions. The RI will serve as a critical testing ground for future technologies such as distributed quantum computing, industrial-grade optical clocks and sensor networks. It will also support advancements in telecommunications, positioning and navigation systems, synchronisation and timing, ensuring that Germany remains at the forefront of global technological innovation. This proposal for a QTF-Backbone aligns well with Germany’s Hightech-Agenda-Deutschland (HTAD), launched in October 2025 by BMFTR~\cite{BMFTRHightechAgenda}, which identifies strategic research fields and quantum technologies as one of six future-defining technologies.

The QTF-Backbone will support ten transformative R\&D areas in quantum technologies and time-frequency metrology. These topics illustrate the infrastructure's potential to drive innovation across disciplines and are summarised below, with detailed discussion in the main text.

\#1. Quantum Communication and Quantum Repeaters

\#2. Wide-Area Network for Quantum Computing

\#3. Synergy between Time \& Frequency Distribution and Quantum Networks

\#4. Precision Measurements in Fundamental Physics

\#5. Infrastructure Monitoring and Seismic Observation through Optical Sensing

\#6. Global Reference Frames for Navigation and Geosciences

\#7. Contributions to a Unified Height System and Gravity Satellite Missions Support

\#8. Enhanced Astronomical Observations through Timekeeping

\#9. Redefinition of the SI Second via Optical Clock Comparisons

\#10. Resilience at Critical Timing Facilities

In addition to the ten R\&D topics outlined above, the QTF-Backbone holds significant potential to support a wide spectrum of future applications beyond those explicitly identified. Its versatile and high-performance infrastructure can serve as an enabling platform for emerging research directions and interdisciplinary innovation in quantum technologies. As scientific and technological needs evolve, the QTF-Backbone can facilitate breakthroughs in areas such as quantum-enhanced environmental monitoring, advanced quantum simulations, and secure data infrastructures. Furthermore, it can foster new collaborations across academic, industrial, and governmental sectors, driving forward the integration of quantum links into critical infrastructure and societal systems. The adaptability and scalability of the QTF-Backbone make it a foundational asset for the continued exploration and realisation of transformative quantum technologies not yet foreseen.

By implementing this dedicated research network in Germany, the QTF-Backbone will secure capabilities to contribute to scientific advancement, economic opportunities and future essential infrastructure. As a hub for national and international cooperation, the QTF-Backbone will contribute significantly to Europe's research landscape, and specifically to the planned pan-European fibre networks, ensuring continued innovation and global competitiveness in the coming decades.

\section{Concept of the QTF-Backbone infrastructure}
\label{sec:concept}

\subsection{Technical Design}

The QTF-Backbone is envisioned as a wide-mesh national fibre-optic core network enabling scientific and industrial R\&D in quantum technologies and precision time-frequency applications. Connecting testbeds via a national backbone ensures reliable and long-term access to advanced infrastructure for researchers and developers. 
By complementing the existing German research network (X-WiN), the QTF-Backbone will establish a separate, technically dedicated fibre-optic infrastructure for quantum technologies, laying the foundation for a next-generation quantum internet. The technical and structural design for the QTF-Backbone proposed here builds on the 10+ years of experience gained from several isolated testbeds in Germany as well as the detailed planning of the design studies carried out as part of the quantum initiative~\cite{Bundesregierung2023} of the Federal Ministry of Research, Technology and Space (BMFTR), previously Federal Ministry for Education and Research (BMBF) and the CLOck NETwork Services - Design Study (CLONETS-DS) project of the EU~\cite{CLONETSDS2025Deliverables}.

We envision, as described in more detail below, that the QTF-Backbone would become increasingly available to users over four phases during a 10-year installation period. Thereafter, the infrastructure would provide access to users across Germany for at least another decade to carry out R\&D in a variety of scientific fields without being affected by the operational requirements of other data communication networks. The DFN e.V. (Registered Association for a German Research Network, in German Deutsches Forschungsnetzes e.V.) is a suitable organisation to commission with the operational tasks of the QTF-Backbone (see section~\ref{sec:PathtoImplementation} for more details), as they have decades of experience providing broad user accessibility for research and education.

\begin{figure}[h]
\centering
\includegraphics[width=0.9\textwidth]{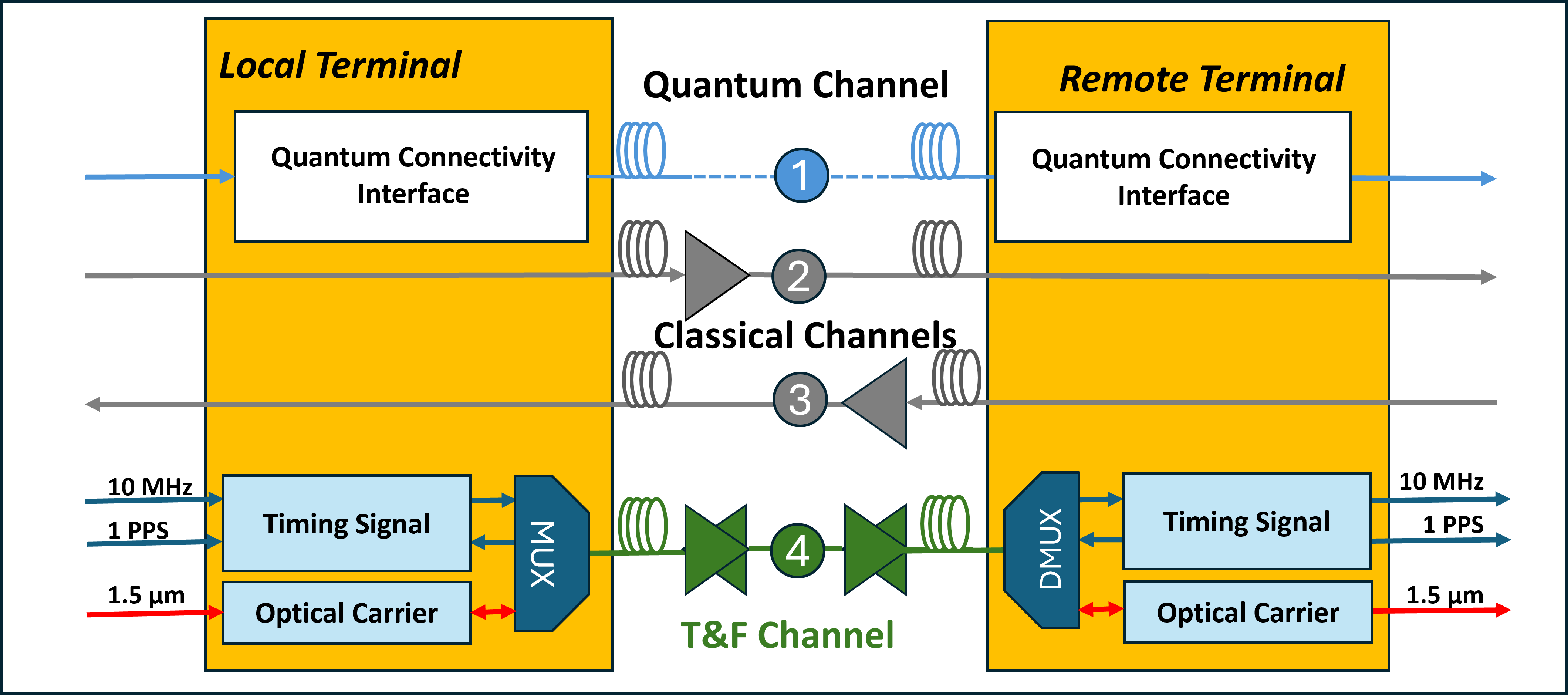}
\caption{Network Layer: Four-fibre (two-pairs) approach connecting two Points of Presence (PoPs) with separate dark fibres for quantum communication (fibre 1) and time and frequency dissemination (fibre 4).}\label{fig1}
\end{figure}

The research infrastructure (RI) will connect distributed users and will consist of three layers:

1. The basic network layer shown in  Figure~\ref{fig1} will comprise dedicated dark fibres (two pairs) and basic network equipment (e.g. amplifiers, switches, routers) to connect key research institutions. The dark fibres will be provisioned commercially, mostly through long-term rental or so-called IRUs (indefeasible right of use) of existing telecommunications infrastructure. In some cases, for example to the BKG Geodetic Obervatory Wettzell, new fibre routes will have to be deployed.

\begin{figure}[h]
\centering
\includegraphics[width=0.9\textwidth]{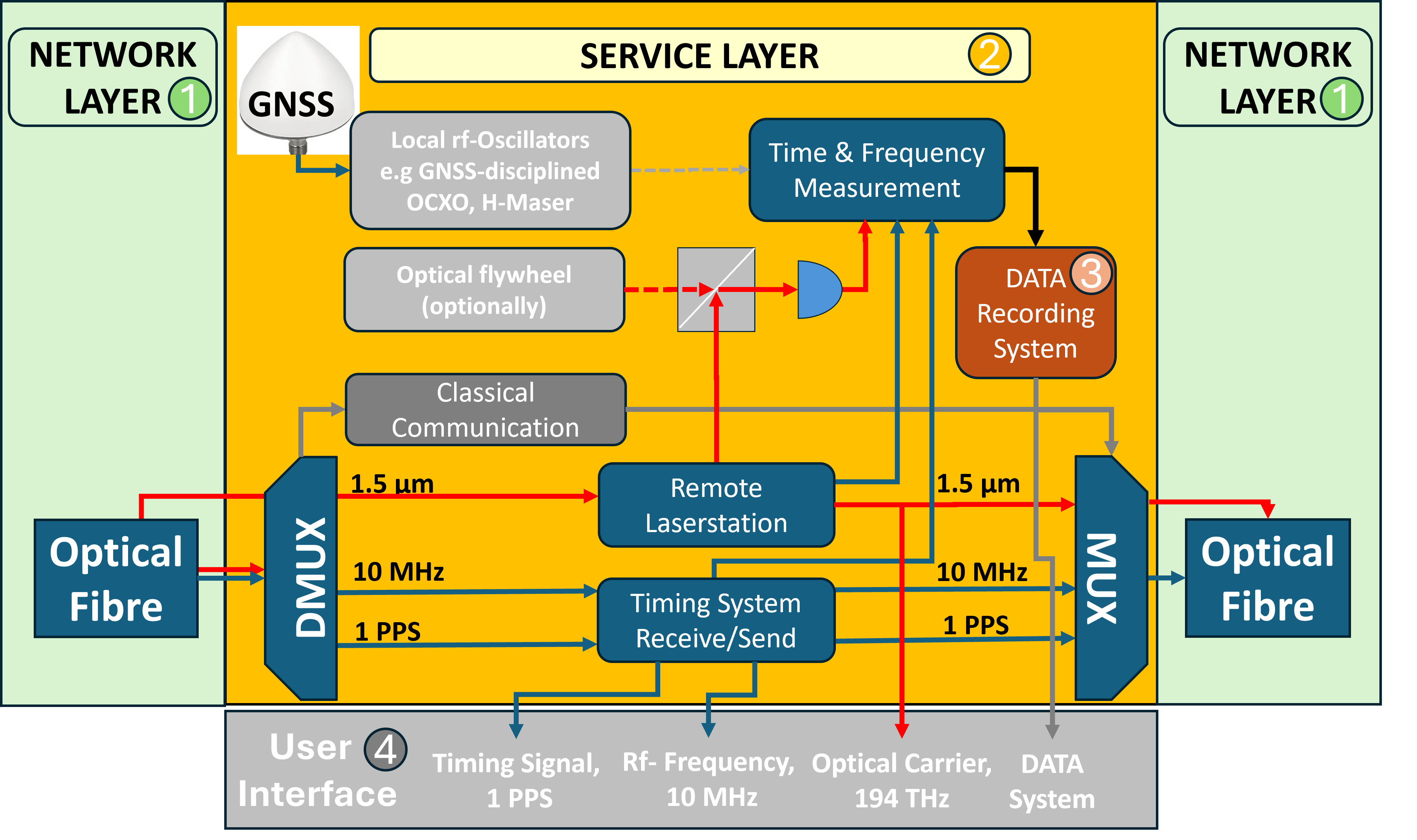}
\caption{Interface of the layers 1, 2, 3 and 4: Typical time and frequency infrastructure of a PoP, e.g. a GNSS-disciplined Oven Controlled Crystal Oscillator (OCXO) or Hydrogen Maser. PoPs will provide a timing infrastructure as well as an optical infrastructure that allows for time or frequency intercomparisons throughout the network.}\label{fig2}
\end{figure}

2. The service layer shown in Figure~\ref{fig2} will provide high-precision T\&F references throughout the network. In the first instance, the backbone will be equipped for distribution of an optical reference at 194.4 THz along with 10 MHz and Pulse Per Second (PPS) reference signals; these signals will be fed in from a source maintaining high-performance T\&F signals such as at the PTB. All installations will be prepared for future multiplexing of other services. The service layer will also provide secure and convenient data channels for management and classical communication as well as a server infrastructure for example to host key management systems for (quantum) cryptographic applications.

3. The data layer shown in Figure~\ref{fig2} will provide a comprehensive IT infrastructure for collecting and storing scientific data and for making it securely accessible to appropriate user groups. Specifically, information collected from the T\&F services is required for applications such as calibration of high-precision measurements but also directly reflects the time-dependent delay of the fibre used for transmission, which is susceptible to environmental factors and can therefore act as a sensor. The data layer will be accessible to scientists world-wide via a repository as described in section~\ref{sec:PathtoImplementation}.

4. This network will be available via a user interface at the connected nodes, or Points of Presence (PoPs), in various interface options for research activities. Interfacing may be realised locally, where user equipment is physically hosted at the PoP, or remotely through user-provided fibre. Some use cases, most prominently in quantum communication, will require physical access to and/or the exclusive use of fibre. Other examples where this might be the case could be research on ultrahigh-bandwidth telecommunications, on techniques for synchronisation at the ps-level or below, or on remote sensing. Optical routing and multiplexing capabilities will be integrated into the basic network layer to improve network efficiency and expand use. We note that although the topology of the network is chosen so that it is partially meshed with a redundancy of signals in case of fibre outages, initially the RI will not offer any network corrections in real-time. However, such upgrades will be actively developed via R\&D on the QTF-Backbone and, with appropriate investments, will be integrated into the operational layers over the 10-year implementation phase. A special type of source PoP, serving as a source for high-performance T\&F signals, which are fed into the network (described in point 2), is not shown explicitly in the figures.

Data channels and reference signals will be available to users at the PoPs, primarily to support users of the dark fibre network but may be further distributed, subject to appropriate service-level agreements. The RI will support third parties planning to distribute reference signals or data channels.

For the dark fibre network to deliver optimum value for scientific research, underground cables should be utilised wherever possible, as the level of environmental disturbance directly impacts the transmission performance. The physical properties of the deployed fibre and its geometry should be known to a high level of detail -- this information is critical for planning experiments and interpreting the results. Both requirements will substantially enhance the value of the QTF-Backbone as a well-characterised RI compared to standard telecommunication networks.

Different types of nodes are required in this network:

a. Point of Presence (PoP): a site connected to the network where services are available, capable of hosting scientific equipment. Here, local and regional users will get access to the QTF-Backbone. The locations of PoPs influence the topology of the RI.

b. In-Line Amplifier Site (ILA-S): a site connected to the dark fibre network where equipment to regenerate and route signals associated with the service layer are hosted. ILA-S are determined by the dark fibre provider, subject to constraints on span length or attenuation. The specific location of the ILA-S will be determined by the procurement process. ILA-S may also host user equipment like trusted nodes and KMS hardware, either commercially available ones or systems under research and development, but only to a limited extent (e.g. in terms of space and energy requirements).

To provide access via the user interfaces (4) to the network layer (1), operate the service layer (2) and collect the information made available through the data layer (3), as shown in Figure~\ref{fig2}, specific hardware is required at the nodes (PoPs or ILA-S). These include:

\begin{itemize}

\item Optical switches, including wavelength-selective switches, suitable for routing signals at the single-photon level, to enable flexibility in setting up quantum networks

\item Equipment to characterise and control polarisation and delay of fibres, to provide quantum channels with known and ideally stable transmission properties, or to add noise in a controlled way

\item Fibre-optic communication equipment (amplifiers, switches, routers), to provide bandwidth for classical communication that can be multiplexed with T\&F signals

\item Regeneration equipment for T\&F signals, to ensure a consistently high signal quality throughout the network

\item Data acquisition hardware, to capture information about the propagation delay of the fibres at a rate suitable for sensing applications

\item Servers capable of hosting key management systems, to support heterogeneous QKD networks.

\end{itemize}

\begin{figure}[h]
\centering
\includegraphics[width=0.9\textwidth]{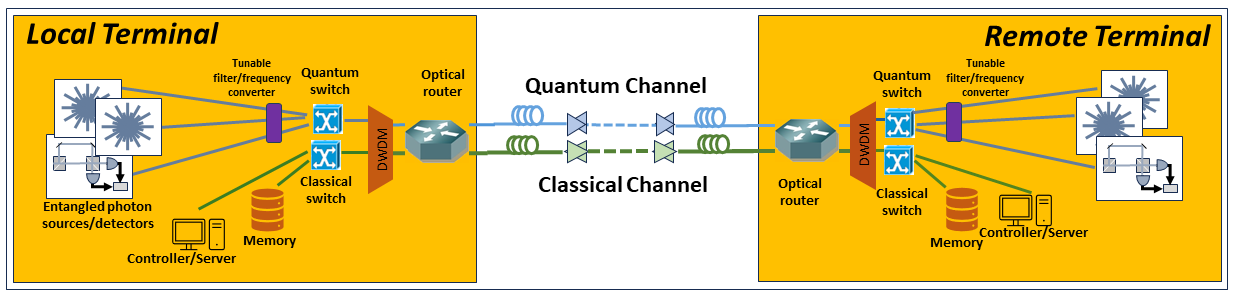}
\caption{Examples of hardware housed in two PoPs or ILA-S for enabling quantum connectivity.}\label{fig3}
\end{figure}

Figure~\ref{fig3} illustrates this typical PoP and ILA-S hardware for enabling quantum connectivity.

Figure~\ref{fig2} and Figure~\ref{fig3} present only examples for components to provide services and to build a PoP or ILA-S. Further details will be worked out in a dedicated technical design phase, which will also involve consultation of prospective users according to the governance of the QTF-Backbone as described in section~\ref{sec:PathtoImplementation}.

Besides their core functionality as part of the network, PoPs must also be able to host scientific equipment for experiments conducted using the RI. Compared to standard telecommunications equipment, these systems impose special demands on the local infrastructure. These include sufficiently large spaces, electricity supply and environmental management (e.g. cryogenic and heat management), but also organisational aspects like access for personnel or regulatory requirements on equipment.

The T\&F distribution techniques required to support services of the QTF-Backbone are mature: commercial off-the-shelf (COTS) equipment is available, albeit from only one or two suppliers. Multiplexing of data and time-frequency signals has been demonstrated many times and is implemented to some extent in most operational networks and links for management purposes. A technical guide written as part of the EMPIR TiFOON project summarises options and potential pitfalls~\cite{turza2022}.

Providing stabilised (with respect to delay and polarisation) fibre for quantum communication purposes requires the accessibility of single-photon-level signals and auxiliary classical signals. This remains an active research area, as explored in BMFTR-funded (previously BMBF-) projects like QR.X, QR.N~\cite{QRN2025}, and QuNET~\cite{QuNET2025}.

For fibre sensing, the first instance use cases will be limited to either data from T\&F services provided through the repository, or physical access to the dark fibre. In the medium term, working with the relevant user communities, we anticipate that additional, specialised services will emerge that, once sufficiently developed, could be integrated into the infrastructure. While the RI will support such development, we will rely on the respective communities to take the initiative. For example, it is conceivable that phase-sensitive Optical Time-Domain Reflectometer (OTDR) or Optical Frequency Domain Reflectometry (OFDR) for seismic monitoring could be integrated. Multiplexed with data and time-frequency signals, it could become part of the suite of services operating continuously across the network and providing data for the data layer.

A comprehensive software solution for monitoring and management of the network will have to be developed. For the data layer, a repository will have to be implemented. Further development will focus on engineering decisions during the planning stage, including topology and equipment selection.

A network operating centre (NOC) will manage maintenance of and access to dark fibres and will monitor basic parameters of the service layer. Where necessary to resolve specific issues it will request assistance from participating institutes, where more in-depth expertise is available. The scietific data and the repository for the data layer will support the ``FAIR'' (Findable, Accessible, Interoperable and Reusable) and ``TRUST'' (Transparency, Responsibility, User focus, Sustainability, and Technology) principles, correspondingly.

\subsection{Performance levels for time and frequency transmission}

\begin{figure}[h]
\centering
\includegraphics[width=0.9\textwidth]{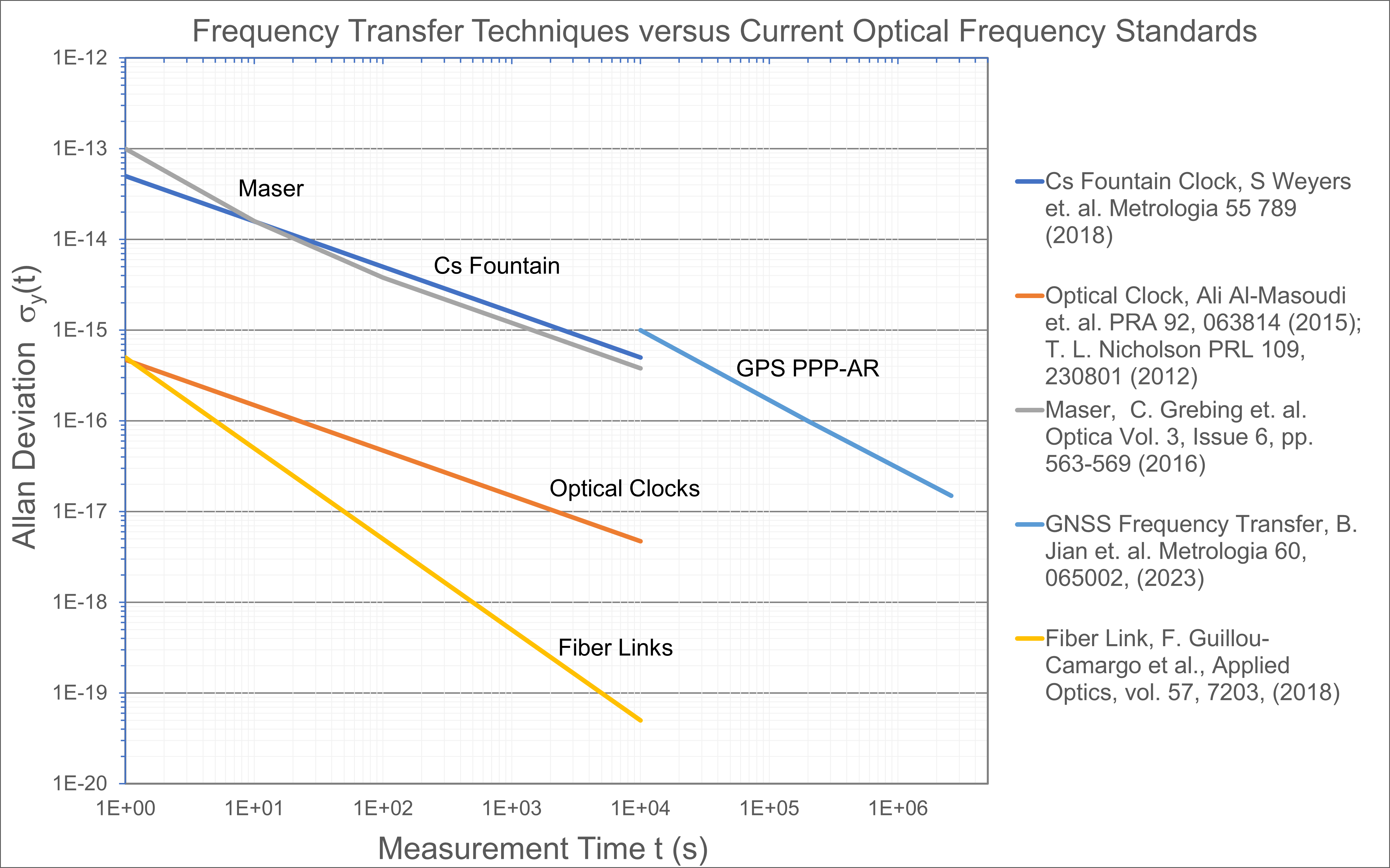}
\caption{This plot (originally shown in CLONETS-DS Deliverable 1~\cite{CLONETSDS2025Deliverables}) shows the instability versus measurement time for representative masers~\cite{Grebing2016}, Cs fountain clocks~\cite{Weyers2018} and optical clocks~\cite{Masoudi2015, Nicholson2012} compared with the frequency-transfer techniques realised via Global Positioning System (GPS) Precise Point Positioning with Ambiguity Resolution (PPP-AR)~\cite{Droste2015, Petit2021} and optical fibre links~\cite{Guillou-Camargo2018}.}\label{fig4}
\end{figure}

A time uncertainty in the sub-nanosecond range and/or a relative frequency uncertainty of better than 1 part in 10$^{15}$ have thus far only been achievable and accessible in national metrology institutes or a few laboratories worldwide. The QTF-Backbone will enable the distribution of such superior T\&F signals to users in industry and research institutions throughout Germany, because this is technically only possible via dedicated bi-directional dark optical fibres, as shown in the figure~\ref{fig4}.

Table 1 gives more technical details on the signals that can be distributed via an optical fibre and compares the performance to GNSS-based techniques. Using these signals, it will be possible to

\begin{itemize}
    \item Compare optical frequency standards (such as optical clocks) to 1 part in 10$^{19}$, two orders of magnitude better than is possible with established techniques like GNSS
    \item Receive ultrastable laser light with an instability below 1 part in 10$^{16}$ (flicker floor), limited only by thermal noise in PTB's world-leading cryogenic silicon cavities
    \item Receive a radio-frequency reference signal with an instability equivalent to an active hydrogen maser, traceable to UTC (Universal Time Coordinated)
    \item Synchronise devices within less than 100 ps across the QTF-Backbone, and within 1 ns to a physical realisation of UTC.
\end{itemize}

The information in the Table~\ref{tab:performance} has been verified in Europe for distances of more than 600 km and can also be expected for the QTF-Backbone.

\begin{longtable}[h]{@{}P{0.25\textwidth} P{0.33\textwidth} P{0.44\textwidth}@{}}
\caption{Approximate performance levels for T\&F transmission using cw optical carrier~\cite{Cantin2021}, ELSTAB~\cite{Krehlik2016, Sliwczynski2019}, WR~\cite{Kaur2022, Dierikx2019}, and, for comparison, GNSS~\cite{Jian2023, Petit2021}. These numbers are provided solely for orientation of what can be expected via the QTF-Backbone RI, as actual performance of any link will depend on its specific parameters. Comparison services are limited by the time and frequency transfer technique while references are additionally limited by the source themselves.}
\label{tab:performance}\\
\hline
\textbf{Service} & \textbf{Signal} & \textbf{Specification} \\

\hline\hline
Optical frequency reference &
CW optical carrier at 1542 nm \newline
Ultrastable cavity stabilised laser, steered to active hydrogen maser, delivered through actively compensated interferometric fibre link. &
\textbf{Instability (ModADEV):} \newline
$<10^{-15}$ @ 1 s (transfer limited, see ``comparison services'' below); \newline
$<10^{-16}$ @ 10 s (transfer and/or cavity limited)\newline
$<10^{-16}$ @ 100 s (cavity limited~\cite{Milner2019}); \newline
$<10^{-16}$ @ 1000 s (cavity limited~\cite{Milner2019}).\newline
$<10^{-15}$ @ $10^4$ s (maser limited~\cite{Schwarz2020}).\newline
\textbf{Accuracy:} $<10^{-16}$ vs UTC(PTB); \newline
\ practically $\lesssim10^{-15}$ vs UTC. \\[0.7em]
\hline
Remote optical frequency comparison &
Mediated by shared optical carrier at 1542 nm, see above. &
\textbf{Instability (ModADEV):} \newline
$<10^{-15}\allowbreak/\tau$ @ 1--1000 s; \newline
$<10^{-18}$ @ $10^{3}$--$10^{5}$ s; \newline
$<10^{-19}$ @ $10^{5}$ s (\(\approx\)1 day).\newline
\textbf{Accuracy:} $<10^{-19}$. \\[0.5em]
\hline
Radio frequency reference &
Radio frequency at 10 MHz, maser steered to UTC(PTB), delivered through ELSTAB. &
\textbf{Instability (ModADEV):} \newline
$<10^{-12}$ @ 1 s; \newline
$<10^{-16}$ @ $10^{5}$ s (\(\approx\)1 day).\newline
\textbf{Accuracy:} $<10^{-16}$ vs UTC(PTB); \newline
\ practically $\lesssim10^{-15}$ vs UTC. \\[0.7em]
\hline
Remote radio frequency comparison &
Mediated by shared radio frequency at 10 MHz, see above &
\textbf{Instability (ModADEV):} \newline
$<10^{-12}$ @ 1 s; \newline
$<10^{-14}$ @ 100 s; \newline
$<10^{-16}$ @ $10^{5}$ s (\(\approx\)1 day).\newline
\textbf{Accuracy:} $<10^{-16}$. \\[0.7em]
\hline
Time reference &
Electrical PPS signal, derived from UTC(PTB), delivered through ELSTAB &
\textbf{Instability (TDEV):} \newline
$<10$ ps @ 1--$10^{6}$ s~\cite{Sliwczynski2019}.\newline
\textbf{Accuracy:} $<100$ ps vs UTC(PTB);\newline 
pratically within a few nanoseconds of UTC. \\[0.7em]
\hline
Time and frequency reference &
PTP-WR (``White Rabbit'') through uni- or bidirectional DWDM network &
Roughly one order of magnitude worse than ELSTAB, see above~\cite{Minetto2023}. \\[0.7em]
\hline
Time and frequency reference &
GNSS Signal in Space, typical GNSS disciplined oscillator &
\textbf{Instability:} \newline
$10^{-11}$ @ 1 s; \newline
$10^{-13}$ @ $10^{5}$ s (\(\approx\)1 day).\newline
\textbf{Accuracy:} frequency $10^{-14}$; GNSS time \(\approx\) 10 ns. \\[0.7em]
\hline
Remote frequency comparison &
GNSS-based techniques for comparison (Integer Precise Point Positioning IPPP, PPP-AR) or TWCP (Two-Way Carrier-Phase satellite time transfer) &
\textbf{Instability:} 
$10^{-14}$ @ $3\times10^{3}$ s (\(\approx\)1 h); \newline
$2\times10^{-17}$ @ $3\times10^{6}$ s (\(\approx\)30 d~\cite{Jian2023}).\newline
\textbf{Accuracy:} $10^{-16}$ @ $10^{6}$ s (\(\approx\)10 d). \newline
Time (Precise Point Positioning PPP or TWSTFT): \(\approx\) 1 ns\\
\botrule
\end{longtable}


\subsection{Integration and unification of existing German and European infrastructures}

In Germany, several states have launched their own quantum initiatives to promote quantum technologies and establish them in the economy through industrial applications. To this end, quantum competence centres have been established in many places, particularly with a focus on quantum communication and quantum computing. At the federal level the quantum technology competence centre (QTZ) has been established at PTB to support the industrial development of quantum technology with an emphasis on metrology and standardisation. PTB also leads the BMFTR-funded (previously BMBF-) project ``Umbrella project for quantum communication in Germany'' (SQuaD), in which an interactive map of testbeds and quantum communication testlinks has been compiled~\cite{SQuaD2025Testbeds}. Several of these testbeds have been implemented in the framework of state initiatives. Table~\ref{tab:competence-centres} gives an overview of the status of the quantum competence centres with a focus on the initiatives for establishing quantum communication testbeds and sub-networks.


\begin{longtable}[h]{@{}P{0.2\textwidth} P{0.75\textwidth}@{}}
\caption{Overview of the status of the quantum competence centres with a focus on the initiatives for establishing quantum communication testbeds and sub-networks. An online interactive map of testbeds throughout Germany is hosted by the BMFTR-funded (previously BMBF-) SQuaD project~\cite{SQuaD2025Testbeds}}
\label{tab:competence-centres}\\
\hline 
\textbf{State} & \textbf{Description} \\
\hline
Baden-Württemberg & QuantumBW strives to develop a state-wide ecosystem for quantum technologies, in particular quantum sensing, quantum computing, and quantum communication and these efforts are complemented by the Center for Integrated Quantum Science and Technology (IQST),which brings together researchers from diverse disciplines across Baden-Wuerttemberg. The total amount for the first period 2023-2027 amounts to more than 100 M€ to foster collaborations between academia and industry to bring quantum devices to the market. Several locations are planning to establish local fibre links and QKD networks, e.g. Stuttgart has demonstrated an inner-city fibre network and will build a multi-node network across various locations to explore networked quantum applications beyond bi-partite communication such as networked communication, secret sharing, or networked computing. \\[0.35em]
\hline
Bavaria & Several local quantum technology centres and high-tech lighthouse projects are pursued in Quantentech Vision Bayern, including the “Munich Quantum Valley” initiative, for which 300 M€ in Bavarian state funds have been earmarked until 2026. Several local quantum networks are being developed and are already in operation in Bavaria, connecting various universities, public research institutes, and industry partners via fibre- and free space links. There are also optical ground stations for satellite-based quantum communication operational. \\[0.35em]
\hline
Berlin & Berlin Quantum initiative funded with 25 M€ by the state of Berlin~\cite{BerlinQuantum} is focusing on fundamental research as well as technology transfer. A tap-proof quantum network is being created on the Charlottenburg campus at the TU Berlin in the joint project tubLAN Q.0 (funded by the BMFTR / previously BMBF). Additionally the Berlin Quantum Communication Testbed offers ca. 34.4 km of dark fibre links and a 3.1 km free space link.   \\[0.35em]
\hline
Brandenburg & Brandenburg’s universities and research institutions have been conducting collaborative, quantum-technology research for many years funded at both the federal and EU levels, establishing strong national and international partnerships. Brandenburg has recognized the strategic importance of key technology and the state has also allocated funding, including 12.8 M€ support from the Brandenburg Future Investment Fund for the Center for Quantum Technology and Applications at DESY’s Zeuthen site~\cite{DESYZeuthenQuantum}. Moreover, the Brandenburg Quantum Technology Network collaborates in partnership with regional, national, and international actors from science and industry to position the capital region as an attractive hub for new quantum technology applications~\cite{BrandenburgRoadmap}. \\[0.35em]
\hline
Hamburg & The Innovation Center Hamburg (IZHH) is part of the German Aerospace Center's (DLR) Quantum Computing Initiative. It serves as a collaborative hub where research institutions, industry partners, and startups converge to develop quantum computing technologies and applications. Located in Hamburg (and in Zeuthen), DESY is a national research center that utilises fibre-optic networks for various scientific applications, including seismic measurements and data transmission for particle physics experiments. \\[0.35em]
\hline
Lower Saxony & The Quantum Valley Lower Saxony (QVLS) initiative was launched in 2020, which supports a range of activities for technology transfer across all Technology Readiness Levels (TRL). The establishment of the Niedersachsen Quantum Link between Hannover and Braunschweig offers the possibility for testing quantum communication and T\&F-distribution on the same fibre pair. The 78 km Niedersachsen Quantum Link is complemented by a 14 km intracity link from the quantum technology competence center (QTZ) at PTB to the High-Tech Startup Hub QVLS-HTI located in the city of Braunschweig. \\[0.35em]
\hline
North Rhine-Westphalia & EIN Quantum NRW is a state-supported network of universities, research institutions, and companies that brings together and promotes scientific and economic interests in the field of quantum technologies in North Rhine-Westphalia. A particular focus is on education and the transfer of basic research into practical applications. North Rhine-Westphalia hosts several testbeds participating in the QR.X, QR.N, PhoQSNET and METRO projects as well as a strong and diverse research landscape in this future-oriented field of quantum communication.  \\[0.35em]
\hline
Saarland & Several institutions received 10 M€ in federal funding for quantum technology projects since 2022. At Saarland University a metropolitan fibre testbed for quantum network demonstrations has been set up in the context of BMFTR-funded (previously BMBF-) projects (Q.link.x, QR.X, QR.N). This testbed will be expanded by further nodes and links in the next years. \\[0.35em]
\hline
Saxony & Saxony is also funding a project for tap-proof data transmission using quantum technology with 8 M€. TU Dresden joined the EU Quantum Internet Alliance project at the end of 2022 and is researching a quantum internet for Europe. \\[0.35em]
\hline
Thuringia & The “Quantum Hub Thuringia” aimed for broad quantum technology development with eleven research institutions from all over Thuringia and a state funding of about 6 M€ over a period of 32 months (ended 2024). Part of the project is extended as part of the “Thuringian Innovation Centre for Quantum Optics and Sensing (InQuoSens)” which started 2024. Furthermore, building up the competences for an application laboratory “Quantum Engineering” in Erfurt was funded by the Thuringian state with 11 M€ which included a fibre testbed between Erfurt and Jena and an optical ground station in Jena. \\
\end{longtable}

QTF-Backbone has the enormous potential to physically interconnect these currently disparate state initiatives and promote synergies e.g. to strengthen the R\&D activities beyond localised activities. A brief description of the existing inter-state initiatives from Fraunhofer-Society and Deutsche Telekom AG are listed in Table~\ref{tab:interstate}.

\begin{table}[h]
\caption{Overview of inter-state initiatives.}
\label{tab:interstate}
\small
\renewcommand{\arraystretch}{1.2}
\begin{tabular}{@{}P{0.25\textwidth} P{0.7\textwidth}@{}}
\toprule
Coordinator & Description \\
\midrule
Fraunhofer-Gesellschaft & The Fraunhofer-Gesellschaft is currently setting up a Fraunhofer Competence Network for Quantum Computing (FNQ) on a cross-state basis and financed by the respective federal states. Berlin, Thuringia and Bavaria have joined forces as part of the Germany-wide, BMFTR-funded (formerly BMBF-funden) QuNET initiative~\cite{QuNET2025} to research key technologies for quantum communication. The QuNET alliance has 5 core institutes and a further 47 partners, including 26 industry partners, working on quantum communication technologies through the QuNET+ projects. \\[0.35em]
Deutsche Telekom AG & Deutsche Telekom AG is investigating quantum computing and quantum communication applications; and is leading the initiative for the development of the European-wide tap-proof infrastructure EuroQCI under the name “Petrus”. Deutsche Telekom AG operates a test track for quantum communication between Bonn and Berlin (DemoQuanDT project~\cite{DemoQuanDT2025}) and the SaSER network, which was previously funded by the BMFTR / previously BMBF. \\
\botrule
\end{tabular}
\end{table}


Beyond synergies with the existing state initatives, the QTF-Backbone will actively pursue interconnections with existing fibre networks in Germany. Table~\ref{tab:german-networks} gives an overview of existing German networks and their relevance to the proposed QTF-Backbone.

\begin{table}[h]
\caption{Existing German networks relevant for the QTF-Backbone.}
\label{tab:german-networks}
\small
\renewcommand{\arraystretch}{1.2}
\begin{tabular}{@{}P{0.15\textwidth} P{0.25\textwidth} P{0.25\textwidth} P{0.3\textwidth}@{}}
\toprule
Name & Website & Site / Countries & Relation to the project \\
\midrule
SaSER & -- & Ring structure throughout Germany & Operator: Deutsche Telekom AG; Focus: Secure IT infrastructure and higher data transfer rates. \\[0.35em]
DemoQuanDT & \url{https://www.forschung-it-sicherheit-kommunikationssysteme.de/projekte/demoquandt} & Berlin to Bonn & Operator: Deutsche Telekom AG; Focus: Industrial use cases of quantum communication. \\[0.35em]
X-WiN & \url{https://www.dfn.de/netz/} & Extensive network in Germany & Operator: DFN e.V.; Focus: Support research institutions in Germany. \\[0.35em]
PTB & \url{https://www.ptb.de/cms/ptb/fachabteilungen/abt4/fb-43/ag-434.html} & Braunschweig to Garching via Straßburg & Operator: PTB; Focus: Operate international links with PTB's T\&F infrastructure. \\[0.35em]
Deutsche Telekom (DTAG) & \url{} & Links connecting Braunschweig, Bremen, Frankfurt am Main, Berlin and Nürnberg & Operator: Deutsche Telekom (DTAG); Focus: optical-fiber time transfer links have been established connecting UTC(PTB) in Braunschweig with facilities of DTAG. These fiber links are used for supervision of the performance of network synchronisation with DTAG~\cite{Sliwczynski2020}. This network was recently expanded with connections to the Deutsche Bahn network. \\[0.35em]
BMI BDBOS & \url{https://www.bdbos.bund.de/EN/TheBDBOS/Aboutus/aboutus_node.html} & Throughout Germany & Focus: “Government Network”, Public safety. \\
\botrule
\end{tabular}
\end{table}

\begin{figure}[h]
\centering
\includegraphics[width=0.9\textwidth]{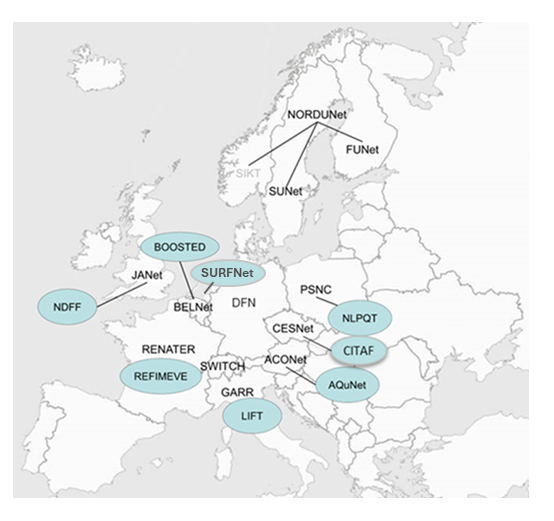}
\caption{Map showing already existing fibre link consortia in Europe (acronyms in blue ovals) that are under development in collaboration with NRENs (acronyms only).}\label{fig5}
\end{figure}

R\&D infrastructures for T\&F and/or quantum communication are already in place in several European countries as shown in Figure~\ref{fig5}. Table~\ref{tab:european-fibre} gives an overview of the  current state-of-the-art achieved in EU countries.

\begin{longtable}[h]{@{}P{0.2\textwidth} P{0.7\textwidth}@{}}
\caption{Summary of the available European optical fibre R\&D infrastructures for T\&F and/or quantum communication as shown symbolically in ~\ref{fig5}.}
\label{tab:european-fibre}\\
\hline
\textbf{Country} & \textbf{Description} \\
\hline
Austria & Austria is in the process of upgrading its nation-wide optical fibre network for the distribution of quantum information and quantum metrology signals which includes establishing new channels and making existing dark fibres accessible. Time-frequency links within Vienna, Vienna-Linz-Salzburg-Innsbruck and Vienna-Brno (Czech Republic) are operational, further connections to REFIMEVE in France and PTB in Germany are planned. \\[0.35em]
\hline
Czech Republic & Czechia went through a major upgrade program from 2020-2023 for their R\&D network to allow the effective use of fibres through FlexWDM systems for dedicated optical channels and services as well as dark channels for precise time, frequency and QKD. The Czech Infrastructure for Time and Frequency (CITAF)~\cite{CITAF2025}  currently comprises more than 2200 km and is scheduled for upgrades in 2025 that will offer time services based on White Rabbit and will cover all major cities using a redundant topology~\cite{Vojtech2022CITAF}. \\[0.35em]
\hline
Finland & Finland has implemented a White Rabbit link to its Very Long Baseline Interferometry (VLBI) station in Kajaani. \\[0.35em]
\hline
France & Since 2021 REFIMEVE is included in France’s National Research Infrastructure (RI) roadmap. France launched REFIMEVE+ in 2012 and its successor T-REFIMEVE in 2024 to provide long-distance frequency transfer and time services to French laboratories. The network spans now more than 2$\times$4500 km and currently provides access to 29 partners~\cite{REFIMEVE2025}, including CERN, with 10 more partners coming online by 2025. \\[0.35em]
\hline
Italy & In Italy the operation of LIFT, also known as the Italian Quantum Backbone (IQB)~\cite{INRIM2025Backbone}, started in 2013 with an 1850 km infrastructure that was at first primarily used  for T\&F dissemination, but is composed of a pair of fibres so that one fibre can be devoted to T\&F distribution and the second fibre for Quantum Communication activities or any tests that investigate coexistence and synergies between the technologies~\cite{Calonico2016ETSI, Calonico2019NuovoSaggiatore}. \\[0.35em]
\hline
Netherlands & SURF, the Dutch NREN, offers a time and frequency service based on White Rabbit and delivered via their fibre optic infrastructure~\cite{SURFTimeFrequency}. Reference signals (10 MHz and 1 PPS) are provided by the Dutch National Metrology Institute, VSL. The White Rabbit signal is currently available at 11 locations in the Netherlands, including the European Space Agency (ESA) in Noordwijk, and has already been used for research into advanced positioning~\cite{Koelemeij2022, Tiberiusnavi2023} and atomic clock~\cite{Boughdachi2026} technologies, for example. SURF are also exploring quantum applications and is part of the Quantum Internet Alliance (QIA)~\cite{QuantumInternetAlliance}. \\[0.35em]
\hline
Poland & Poland started its T\&F service in 2012 and currently operates an infrastructure covering 2400 km with uni-directional DWDM systems extending into Lithuania~\cite{Vojtech2023}. As part of the NLPQT (National Laboratory for Photonics and Quantum Technologies)~\cite{FUW2025}, combined optical carrier, 10 MHz and PPS distribution will be rolled out in a fibre network linking several major hubs of photonic technologies R\&D. \\[0.35em]
\hline
Switzerland & Switzerland has implemented operational fibre links between METAS (Bern), the Universities of Basel and Zurich. Connections to CERN (Geneva), Neuch\^{a}tel, and Milano are being discussed. Fibre links disseminate optical as well as radio frequency and time references and have been used for seismic detection~\cite{Husmann2021}. \\[0.35em]
\hline
United Kingdom & In the UK fibre-based services for time and frequency distribution are put in place to eliminate reliance on GPS/GNSS as spoofing and jamming attacks are massively increasing (National Timing Centre (NTC) programme)~\cite{NPL2025NTC}.  The National Dark Fibre Facility (NDFF) in the UK also supports the fibre-based UK Quantum Network~\cite{QCommHub2025WP1}  that connects the metro-scale networks in Bristol and Cambridge. Recently, a fibre link connecting the University of Birmingham to the National Physical Laboratory (NPL) and supporting multiple time and frequency services was presented. NPL is also connected to the European fibre network for optical clock comparisons. \\
\end{longtable}


Europe is in a globally unique position regarding the number and density of optical atomic clocks being developed, often at national metrology labs. PTB is one of the major contributors to these efforts. Accordingly, proposals for pan-European backbone structures supporting optical T\&F distribution via fibre networks have also been made in other national and European projects. Examples of this are the EU project CLONETS-DS~\cite{CLONETSDS2025} and the bid for ESFRI status, FOREST (Fibre-based Optical netwoRk for European Science and Technology). Furthermore, GÉANT, the collaboration of the European National Research and Education Networks, has established a task force that strives to link those labs by optical fibres, thus creating a network of optical clocks. Though still in its infancy, this vision is highly regarded internationally. Connecting the QTF-Backbone to this European network would give German R\&D users access at the European level.

We outline here some of the most current developments of these efforts to establish a long-term GÉANT Core Time/Frequency Network (C-TFN) comprising federated cross-border T\&F links that interconnect existing national T\&F networks~\cite{Zenodo2024CLONETSDS}. In September 2024 a pilot link between the Pozna\'{n} Supercomputing and Networking Centre (PSNC) in Poland and the PTB in Braunschweig was put in place by GÉANT as a first step to the GÉANT Core Time/Frequency Network (C-TFN). Within the current GÉANT project phase GN5-2, procurement is underway to extend the pilot link through the Netherlands and Belgium all the way to France. A second link spanning Italy, Austria, Germany, Czechia, and Poland is in preparation. GÉANT will also continue its investigations into Quantum Communication; a first Long-Distance QKD trial was conducted by Toshiba Europe, GÉANT, PSNC in Poland, and Anglia Ruskin University (UK) to demonstrate QKD within commercial telecommunication networks~\cite{GEANT2024QKDdemo, GEANT2025Event1675}. This experiment was conducted over a 254 km telecom network testbed provided by GÉANT, spanning from Frankfurt to Kehl.

Additionally, the QTF-Backbone’s integration with European research infrastructures also presents significant opportunities for collaboration with CERN, the world’s leading laboratory for particle physics. CERN’s ongoing development of ultra-stable time and frequency distribution—such as its connection to the REFIMEVE+ network~\cite{CERN2025OpticalFibreLink, CERN2025QuantumCommunicationNetworks} and the deployment of the White Rabbit protocol for sub-nanosecond synchronisation~\cite{CERN2025WhiteRabbitCollabLaunch, CERN2025WhiteRabbitQuantumEntanglement, Stolk2024} aligns closely with the QTF-Backbone’s goals. By interconnecting with CERN’s fibre-optic infrastructure, the QTF-Backbone could enable cross-border distribution of high-precision timing signals and support joint initiatives in quantum communication, such as Quantum Key Distribution (QKD) and secure data transfer\cite{Meda2025QKDTimeDissemination}. This synergy would not only enhance CERN’s experimental capabilities, including antimatter research and accelerator synchronisation, but also strengthen the QTF-Backbone’s role as a pan-European hub for advanced metrology and quantum technologies.

\section{Anticipated Impact of the QTF-Backbone on R\&D}
\label{sec:RandD}

In this section we present summaries of several R\&D topics that we the authors anticipate would be enabled and significantly advanced by the realisation of the QTF-Backbone infrastructure. For example, the QTF-Backbone would support the integration of quantum computing nodes into a broader network, facilitating distributed quantum computing and enabling operational in-field tests of quantum entanglement over large distances. This infrastructure could also enable breakthroughs in quantum sensing and metrology and accelerate the path toward a functional quantum internet. Furthermore, the potential of a QTF-Backbone as a new reference system for applications with the highest demands on time or frequency signals has already been shown to be necessary and useful for a range of R\&D topics as described here within. All of the R\&D topics enabled by the QTF-Backbone are transformative, fostering technological innovations with broad scientific and societal relevance.

\subsection{Transformative R\&D topic \#1: Quantum Communication}

Quantum communication R\&D plays a key role in advancing technologies in quantum cryptography, quantum computing, and quantum entanglement. Several institutions and research projects throughout Europe are leading efforts in these areas. R\&D addresses QKD, quantum repeaters, and entanglement distribution across networks. The QTF-Backbone would provide an essential infrastructure for testing and scaling these technologies as well as future implementations of quantum communication.

A key R\&D challenge in enabling future quantum communication is the development of secure networks in complex topologies and over long-distance networks. Current networks, like those in the first developments in the EuroQCI~\cite{EuroQCI} initiative, mainly offer point-to-point security due to ``measure and forward'' techniques, which limits the range for end-to-end security, at reasonable key rates for technology and protocol agnostic quantum communication, between 100~km and up to 200~km. Expanding these networks is essential for developing and testing advanced implementations of quantum cryptographic protocols. These include QKD with continuous and discrete variables (CV-QKD and DV-QKD), deterministic single-photon sources, measurement-device-independent QKD or even fully device-independent QKD (MDI/DI-QKD), and entangled-photon-based protocols~\cite{Pirandola2020}. The network also provides the possibility for developing and testing cryptographic building blocks beyond QKD and in distrustful network scenarios~\cite{Broadbent2016, Bozzio2024, Vajner2024}. Examples are quantum coin flipping, quantum bit commitment, and quantum oblivious transfer building the basis for securing certification and authentification schemes in the quantum realm.

A major hurdle for realising quantum cryptographic tasks in an operational fibre network connecting distant locations, is attenuation of light and the deterioration of quantum states in optical fibres. Quantum repeaters~\cite{Avis2023}, which can relay entangled states without disturbing them, are needed to extend communication across several hundreds or even thousands of kilometres; before reaching that goal, the implementation of trusted nodes is required. Satellite QKD requires local networks to be integrated into existing ground infrastructure. Research exploring and testing these technologies in actual use cases requires dedicated fibre networks. Another challenge is distributing quantum entanglement across multiple nodes for secure communication and future quantum internet applications. Experiments on entanglement swapping and synchronisation are essential for scaling up these networks.

For example, a nationwide fibre network would facilitate the distribution of quantum entanglement between multiple cities, a key step in developing applications such as distributed quantum computing and sensor networks. By testing entanglement over large distances, researchers could explore the scalability of these technologies for practical applications. CV-QKD could also be tested at scale, providing valuable insights into its performance and scalability under practical conditions. The network would enable researchers to assess how CV-QKD protocols function over long distances, which is crucial for validating their use in secure communications.

Furthermore, the fibre network would provide a platform to experiment with complex quantum-network topologies, such as mesh or multi-node networks, which are necessary for scaling up quantum-internet applications. This would allow for the exploration of how quantum cryptographic protocols and building blocks behave in more intricate network setups. Lastly, the network would serve as a testing ground for quantum cryptographic systems under operational conditions, allowing researchers to address practical issues like signal loss, interference, and potential attacks. This will ultimately pave the way for field-deployable, resilient, and ultra-secure quantum communication networks.

\subsection{Transformative R\&D topic \#2: Beyond QKD}

Entanglement describes non-local correlations beyond classical causality. Entanglement distribution is associated with high-fidelity transport of quantum information over short and long distances where the fragile phase of quantum superpositions at the single or multi particle level plays a central role, e.g. in developing quantum memories~\cite{Lvovsky2009}, a requirement for realising quantum repeaters, and establishing long-reach quantum communication~\cite{Sangouard2011}. The German research community has been very active and collaborative on this topic for many years and their current approach to building elementary modules for quantum repeater systems is detailed within a white paper~\cite{vanLoock2020}. Furthermore, from a fundamental point of view, active research is pursued e.g. to test the violation of Bell inequalities in comprehensive ways~\cite{Hensen2015, Liu2018}, or to check against loopholes in practical implementations~\cite{Salart2008, Valivarthi2016}. Current progress towards experimental implementation of quantum-network technologies in laboratory environments and on fibre testbeds include creation of entanglement of remote quantum memories over fibre links~\cite{vanLeent2022, Stolk2024, Knaut2024, Hanni2025} including entanglement swapping operations~\cite{Krutyanskiy2023} and teleportation from an atomic qubit to a telecom photon over a metropolitan fibre link~\cite{Kucera2024}.

Entanglement is not just a cornerstone of quantum communication --- it is emerging as a powerful resource for enhancing classical communication networks as well~\cite{Bassoli2021}. A well-known example is the superdense coding protocol, where shared entanglement allows the transmission of two classical bits by sending just a single qubit via a dedicated quantum channel. This effectively doubles the classical communication rate and highlights the potential of quantum resources in classical settings. Beyond superdense coding, recent research --- both theoretical and in simulated environments --- explores how an additional, entanglement-powered layer can be integrated into classical network architectures. This quantum-assisted layer enables several advanced functionalities that enhance performance, security, and reliability across the network stack. For instance, entanglement can help mitigate denial-of-service (DoS) attacks by enabling authentication directly at the physical layer, filtering out malicious traffic before it reaches higher-level systems. It can also enable physical-layer security mechanisms without relying on QKD, offering tamper-resistance and signal integrity through entanglement correlations alone. Moreover, entanglement can improve the performance of semantic and goal-oriented communication --- where the goal is not just to transmit data, but to achieve specific tasks more efficiently and communicate meaning, as opposed to the transmission of partial information. This is particularly relevant in AI-driven or edge computing scenarios, where rapid, low-latency decision-making is essential. Authentication protocols also stand to benefit, with entanglement enabling fast, lightweight, and highly secure identity verification methods that go beyond classical cryptography. Overall, leveraging entanglement in classical networks opens up new ways to reduce latency, increase trustworthiness, and significantly boost security. It is not about replacing classical communication but about upgrading it with quantum-native capabilities that reshape what is possible in future networks.

Entanglement distribution can also be leveraged to establish a wide-area network for quantum computing. Here we describe an exemplary application. Commercial applications often offload computation to external servers, raising concerns over data privacy, for example, because of the use of untrusted hardware. Encryption and authentication can be used to protect distributed calculations. Another approach is blind quantum computation (BQC), which provides a way for a client to execute a quantum computation using one or more remote quantum servers while keeping the structure of the computation hidden~\cite{Broadbent2009} as demonstrated in 2012~\cite{Barz2012}. The BQC application was thereafter further verified~\cite{Barz2013, Greganti2016}, reviewed in 2017~\cite{Fitzsimons2017}, and recently demonstrated using matter-based qubits~\cite{Wei2025, Main2025}. Elaborate protocols like BQC require the distribution of complex resource states via a network and the synchronisation over multiple to many nodes,  which poses additional network requirements. Synchronisation of servers via high-performance optical T\&F distribution via fibre networks can support the efforts to test and implement such protocols and make the computation private over longer periods.

In all such applications, timing plays a crucial role as soon as in-field implementation is concerned: it enables the synchronisation between distant nodes (e.g. ~\cite{Stolk2024, Liu2024}), ensures a common time-base between remote time-tagging modules~\cite{Valivarthi2016} as well as stable visibility of coincidence counts between interfering photons that travelled via connecting fibres. To verify and employ the distributed entanglement resource in a quantum network, furthermore a common frequency reference frame is required. In some foundational tests, the accuracy of timing is critical to put limits on superluminal influences that could explain entanglement in the framework of existing theories~\cite{Salart2008, Valivarthi2016}.

\subsection{Transformative R\&D topic \#3: Synergy between Time \& Frequency Distribution and Quantum Communication}

Major investments in Europe and internationally are already being made to establish quantum communication fibre networks for transferring quantum-secure information and to prepare what is generally known as the quantum internet. Many of the techniques employed for quantum communication critically depend on time or phase synchronisation.  The integration of T\&F signals with quantum communication requires further investigation to realise its full potential.

For example, critical infrastructure such as distributed power systems for energy production and distribution, can directly benefit from simultaneous quantum communication to secure information exchange, and precise timing to serve as a substitute for GNSS signals for synchronisation. This can bolster data integrity and confidentiality, safeguarding critical information and enhancing the systems's resilience against potential threats.

The work of C. Clivati and co-workers in Italy~\cite{Clivati2022NC} demonstrated impressively that clock network services not only can, but should, coexist with quantum communication networks in order to realise operational quantum communication. With path-length stabilisation techniques similar to established techniques for optical T\&F distribution via fibre networks being employed, the quantum-bit-error-rate due to channel length variations were shown in their work to have been reduced to less than 1\%.

The QTF-Backbone would provide several clear advantages for quantum communication network techniques in need of local and distributed precision frequency control. Furthermore, field testing of novel methods will be greatly simplified by the availability of synchronised local clock bases, which eliminates the need to develop proprietary synchronisation mechanisms at early stages of development. Additionally, the availability of synchronised local clock bases with the QTF-Backbone will enable the development, testing and implementation of protocols exploiting precise timing across multiple nodes. We anticipate that the development cycle of innovative techniques for quantum network components and strategies can be substantially accelerated through the QTF-Backbone, especially if technologies deployed in the QTF-Backbone are later also adopted for commercial products.

\subsection{Transformative R\&D topic \#4: Precision Measurements in Fundamental Physics}

It is widely recognised that precision spectroscopy of atoms and molecules provides an important approach for testing the laws of physics as they are formulated today. Examples are Lorentz Invariance~\cite{Dreissen2022, Sanner2019}, Charge, Parity, and Time reversal (CPT) symmetry~\cite{Akbari2026}, and Local Position Invariance~\cite{Lange2021, Filzinger2023}. Furthermore, the most accurate theory of physics, Quantum Electrodynamics, can be tested~\cite{Blaum2021, Morgner2023}. Moreover, spectroscopy of atoms and molecules is key to providing input data for establishing accurate values of the fundamental constants. A prime example of research on the precision spectroscopy of atoms is the laser spectroscopy of hydrogen atoms performed under the lead of the Max-Planck-Institute for Quantum Optics (MPQ) in Germany. It led to the most precisely determined fundamental constant, the Rydberg constant\cite{Grinin2020, Mohr2024}, insights on the charge radii of protons~\cite{Pohl2010} and deuterons, and on lepton universality. The studies are also key for enabling tests of CPT symmetry by comparing transition frequencies of anti-hydrogen with those of hydrogen~\cite{Baker2025}.

Frequency measurements in laboratories outside of NMIs are challenging. They require the laboratories to individually obtain and maintain expensive references, and still, the accuracy of the measurements is often limited by the references  themselves (e.g. hydrogen maser, caesium or rubidium clock). Today, activities in precision spectroscopy of atoms and molecules in non-metrology institutes are severely hampered by this situation. The measurements performed at MPQ also provide a case study for the benefit of a long-distance T\&F link between PTB and an external institute. To further improve the accuracy of the measurements of the 1S-2S transition in hydrogen, a more accurate reference than the available hydrogen maser and caesium clock at the institute were necessary. This was achieved in a ground-breaking experiment, in which a caesium fountain clock signal was transferred from PTB to MPQ via a 920 km fibre link~\cite{Matveev2013}.

A variety of spectroscopy projects are in progress at non-metrology institutes in Germany (on molecular ions, highly-charged ions, neutral atoms). They would greatly benefit from the availability of optical carrier waves with either ultrastable frequency or ultra-accurate frequency, or both. This would eliminate the need for local ultrastable optical cavities with 10$^{-16}$ level performance.

Beyond supporting local projects, the QTF-Backbone's interconnection of laboratories would also enable the establishment (and, on the European scale, expansion) of a network of optical clocks as a distributed sensor for physics beyond the Standard Model. In this network, the frequencies of the clocks can be compared in real time, making it possible to search for signatures of tiny oscillations of various fundamental constants, an effect that could arise because of the existence of Dark Matter.

Many other fields and applications would benefit from QTF-Backbone's distributed general reference frequencies although they will not require the ultimate performance at 10$^{-18}$ levels but rather will need availability at a level less than the current realisation of the SI unit, i.e. 10$^{-16}$ level. These include lower-level applications such as laser-frequency calibration, length interferometry, sensing of atmospheric and greenhouse gases, and remote wavelength-standard calibration or synchronisation and timing of accelerator facilities.

\subsection{Transformative R\&D topic \#5: Infrastructure Monitoring and Seismic Observation through Optical Sensing}

Distributed Fibre Optic Sensing (DFOS) is a series of well-established techniques for turning a fibre optic cable into distributed, spatially resolved sensors for dynamic strain, temperature and/or strain~\cite{Jousset2018, Currenti2023, Marra2018EarthquakeInterferometry}. Dynamic strain and temperature can be spatially resolved along a single continuous fibre strand with a resolution at the meter scale~\cite{Jousset2018}. Distributed dynamic strain sensing (also called DAS, distributed sensing) allows measurements with amplitudes as low as 10$^{-9}$ m/m/s (nanostrain/s) within fibre strand even on telecom cable~\cite{Currenti2023}, allowing earthquake to be accurately observed in faults~\cite{Jousset2018} and at volcanoes~\cite{Jousset2025FiberVolcano}. Due to its reliance on backscattering, DAS has a limited range of the order of 100~km. More recently, the focus has turned on methods working in transmission, enabling much longer ranges. The challenge here is overcoming the lack of spatial resolution. Both approaches based either on backscattering or transmission, aim to extract environmental parameters and identify event patterns. For example, based on the observation of a magnitude 3.9 earthquake in south‐eastern France, Noe et al.~\cite{Noe2023, Mueller2024} demonstrate how fibre sensing, using by-product data from an optical frequency transfer system, can be used in earthquake detection and characterisation.

The QTF-Backbone, a dark-fiber network, enables the simultaneous deployment and comparison of various techniques -- such as transmission and backscattering -- while the interpretation of results is streamlined by precise knowledge of the fiber’s location and environmental conditions, both of which are provided by the QTF-Backbone infrastructure. Information about the time-dependent phase delay and polarisation-state is given as a by-product of time-frequency and data services and will be available through the data layer. Having a whole network of fibre links available is expected to open new possibilities for locating sources of seismic activity. The flexibility of an infrastructure fully devoted to scientific research allows modifications to the hardware, including that for operating the services, which would be inconceivable in a commercial production environment.

\subsection{Transformative R\&D topic \#6: Global Reference Frames for Navigation and Geosciences}

Applications like satellite navigation and monitoring of global change processes depend on accurate, stable global reference frames. The maintenance of these reference frames is a continuous national and international task and is realised using space geodetic techniques like Very Long Baseline Interferometry (VLBI), Satellite/Laser Laser Ranging (SLR/LLR), GNSS and Doppler Orbitography and Radiopositioning Integrated by Satellite (DORIS). All these techniques are based on precise measurements of signal travel times and thus rely on locally operated stable clocks. An integral part of the position determination task for the one-way measurement techniques VLBI, GNSS, DORIS is the precise synchronisation of the involved clocks, thus degrading the degree of freedom. A precise external synchronisation of the involved clocks has a high potential to increase the stability of the geodetic solutions and position estimates and to reduce the contamination by unmodelled atmospheric delays.  The additional estimation of clock parameters could be completely avoided if a coherent time scale with picosecond accuracy is realised between the different observatories as well as GNSS reference stations, thus stabilising geodetic solutions. A precise T\&F link between the observatories will enable improved accuracy in space geodesy and the development of new observation and data-analysis strategies.

For example, with a recent installment steps have been taken to realise an optical T\&F distribution system with active real-time compensation at the picosecond level at the BKG Geodetic Observatory Wettzell to identify inter- and intra-system biases~\cite{Kodet2025}. In order to exploit the full advantage of this new technique, experiments together with other VLBI stations like Effelsberg and, perspectively, with stations of the International VLBI Service for Astrometry and Geodesy~\cite{VLBI2025Website}, are to be done. Similarly, GNSS receivers shall be installed at nodes of the QTF-Backbone to demonstrate the performance improvement expected in large geodetic networks by running the receivers on a common clock.

\subsection{Transformative R\&D topic \#7: Contributions to a Unified Height System and Gravity Satellite Mission Support}

As of today, European countries, even neighboring countries, do not have a unified height system based on the same reference level. The accuracy and reliability of height determination suffer from progressive error accumulation in classical spirit levelling and network distortions. To eliminate these problems, the International Association of Geodesy (IAG) is establishing the International Height Reference Frame (IHRF)~\cite{GGOS2024, Sanchez2021, Sanchez2024} as the foundation of a unified global height system. In the IHRF, the height coordinate will be represented by a geopotential number -- and only clocks can provide direct measurements of those geopotential values by measuring the differences in gravity potential. The QTF-Backbone will provide a first-order network of reference sites where optical atomic clocks can be compared to high accuracy, as required for chronometric levelling (Relativistic Geodesy) in order to determine height and gravity potential differences~\cite{Grotti2024, Mueller2018}.

Furthermore, capturing the variable part of the gravity field is a key element to understand the underlying change processes and mass transport processes in the Earth system~\cite{Vincent2024}. Many of these large-scale mass transport phenomena (ground water levels, rise of sea water level, melting of ice sheets) are related to climate change. Via use of the signals provided by the QTF-Backbone between ultra-accurate clocks, satellite gravity missions that determine the spatio-temporal variations of the gravity field could be validated with a sufficiently dense grid of ground truth measurements of the geopotential. The local height systems are not able to provide a suitable reference for international or global studies or projects on a large scale. The main requirement in this regard is the establishment of a global physical height system that allows all heights around the world to be referenced to the same reference surface (vertical datum). The establishment of an accurate, consistent, and well-defined global height system would have numerous positive impacts, including:
\begin{itemize}

\item 
Providing a consistent and accurate reference for the connection of national and regional height systems;

\item
Eliminating inconsistencies in gravity anomalies and height information arising from the use of different reference (zero-height) levels;

\item
Enabling the combination of geodetic levelling and oceanographic techniques to determine the sea surface over long distances;

\item
Supporting studies of global change, including climate change monitoring, mean sea-level variations, instantaneous sea-surface modelling, polar ice-cap volume estimation, post-glacial rebound investigations, and land-subsidence analyses;

\item
Providing a reliable reference frame for the consistent analysis and modelling of global phenomena governed by the Earth’s gravity field, such as mass redistribution in the oceans, continents, and Earth’s interior, as well as global ocean circulation and other geophysical fluid processes;

\item
Facilitating the precise combination of physical and geometric height information, thereby maximising the benefits of satellite geodesy (e.g. the integration of satellite positioning with gravity-field models for a globally unified, high-precision height determination).

\end{itemize}

For example, non-tidal gravity field variations of the Earth are monitored by the satellite gravity mission GRACE-FO~\cite{Kornfeld2019}. While the satellite mission and successors yield spatial gravity potential variations, point-wise measurements from clock comparisons can help validate and supplement satellite data. Since the interest here lies in the time-variable geopotential, constant offsets of the frequencies of the clocks are irrelevant. Long-term observations (on the scale of decades) aim to quantify the temporal evolution of the gravity field on spatial scales of a few 100~km. The QTF-Backbone proposed here, could fulfill these requirements and enable Germany to contribute internationally to a unified height system and enhance the results of the gravity satellite missions.

The BKG Geodetic Observatory Wettzell is Germany's space geodetic fundamental station. It comprises all relevant measurement techniques of space geodesy at one site.  A tie between different techniques is implemented, that is indispensable for establishing the consistency of the geodetic parameter set. One essential component of the observatory will be a highly-accurate local optical clock, so that clock frequency can be introduced as a novel tie into the arsenal of methods of space geodesy. It will be important to compare the local clock with the clocks at PTB via the QTF-Backbone, in order to determine the geodetic height and to precisely monitor the local clock performance. Furthermore, the connection to the backbone will implement direct comparisons with clocks at other European geodetic observatories. These observatories are well suited to provide a unique interface between high-accuracy ground clocks and clocks orbiting  in space, a requirement for a uniform global height system~\cite{Vincent2025, Vincent2024ClockSeaLevel}.

\subsection{Transformative R\&D topic \#8: Enhanced Astronomical Observations through Timekeeping}

The VLBI observing method is based on measuring the group delay of signals emitted by remote astronomical sources, like quasars. From the observed time delay between the telescopes of the observing stations, the pointing direction towards the respective quasar and images with high angular resolution are obtained. During the observation, the telescope timescales are not synchronised. By sacrificing some of the observations, the clock offsets from each station with respect to an arbitrarily chosen reference station are determined. To remove systematic errors, VLBI processing infers piecewise linear ``clock corrections'' for every 2-hour interval of observation. These corrections combine, in an unknown way, errors due to the lack of coherence of the involved clocks in the widely separated stations, and to the limiting coherence of the atmospheric propagation delay. With the increased bandwidth of the VLBI Global Observing System (VGOS) of 2.2 to 14 GHz, instead of the current S- and X- band windows, there is already demand for improvement. Significant improvements would be achieved if phase delay measurements could be realised between distant sites. Fibre links are currently the only available method for reaching the performance levels needed in the future.

Radio telescopes and interferometers, including VLBI, are used to capture high-resolution images of astronomical sources. These systems require high-stability frequency references, which the QTF-Backbone could provide. A compelling near-term application of the QTF-Backbone is VLBI-based astrometry, which deals with high-precision angular position measurements of celestial sources, especially distant quasars that define the International Celestial Reference Frame (ICRF).  The QTF-Backbone would support VLBI-based astrometry by enabling optical fibre links between European VLBI sites~\cite{Clivati2020} and the relevant German institutions: the Max Planck Institute for Radio Astronomy in Effelsberg, the GFZ in Potsdam, and the BKG Geodetic Observatory in Wettzell.

\subsection{Transformative R\&D topic \#9: Enable Innovation of Optical Clocks and a Future  Redefinition of the SI Second}

Optical clock research and the redefinition of the SI-unit ``second''~\cite{Dorflinger2024} are a driving force of T\&F links between remote laboratories. Since optical clocks based on various species of atoms (e.g. neutral strontium, ytterbium-ion, and many more) are still being researched worldwide, it is critical to analyse their behaviour at the highest level via clock comparisons. For non-metrological and smaller-scale laboratories it is challenging to establish and maintain sufficient know-how and facilities required not only to develop optical clocks but furthermore, to carry out these necessary comparisons between optical clocks.

Currently, the best optical clocks~\cite{Aeppli2024, Brewer2019} reach uncertainties better than 10$^{-18}$. The validation of their performance requires identifying systematic shifts, assessing environmental sensitivities, and comparing with independent clocks. Usually this is done by means of a second, locally available clock. As described in R$\&$D topic $\#$4, this may not be possible for a non-metrological institute; Additionally, good practice requires verification against an independent clock, preferably in another institute or location. This demands clock comparisons where transfer noise is negligible relative to the systematic uncertainties of the clocks. This means that the noise contributions of the comparison technique should be at least a factor of ten below that of the clocks, i.e. \(<\)10$^{-19}$.

Even higher relative precision is expected with higher transition frequencies. Recent research by many research institutions throughout Germany and worldwide, beyond NMIs, aims to extend the operation frequency of future clocks into ultraviolet and soft x-ray domains. Potential novel candidates are based on highly charged ions~\cite{King2022} or thorium nuclei~\cite{Tiedau2024}. Development of such clocks, their characterisation and validation can only be achieved via frequency transfer techniques with optical fibres. The QTF-Backbone can meet these demanding needs and ensure that optical clocks will continue to be researched and developed not just by NMIs but the wider scientific community. In addition, the QTF-Backbone will provide connectivity to international T\&F networks like the one being implemented by GÉANT and the planned European RI ``FOREST'', ensuring Germany's optical clock signals are shared with Europe and vice versa, enabling comparisons between NMIs vital for the redefinition of the second and securing long-term access to the most advanced optical clocks in Europe for the German scientific community.

\subsection{Transformative R\&D topic \#10: Resilience at Critical Timing Facilities}

As of today ubiquitous access to high-performance T\&F is provided only in the microwave domain by GNSS, such as GPS, GLONASS, Galileo, or BeiDou. Optical T\&F distribution via fibre network, however, outperform satellite methods by orders of magnitude over continental distances (See Figure~\ref{fig4} and Table~\ref{tab:performance}: Approximate performance levels for T\&F transmission). Fibre-optic networks can therefore provide significantly reduced measurement times, unprecedented uncertainty~\cite{Droste2015}, and in turn resilience and improvements of GNSS. To realise these benefits, with huge medium to long-term societal impact including for future Position, Navigation and Timing (PNT) services, an infrastructure is needed for R\&D to develop and establish the future dissemination of time with enhanced performance.

The proposed QTF-Backbone can contribute to an improvement of the well-established T\&F transfer techniques based on satellites. One established satellite time transfer technique is Two-Way Satellite Time and Frequency Transfer (TWSTFT), which relies on the path reciprocity of the transmitted signals~\cite{Banerjee2023}. TWSTFT, with its signals typically in the Ku band, is a technology that is completely independent of GNSS and is operated by a number of institutes that continuously contribute to UTC~\cite{Jiang2018, Jiang2019}. However, it is well known that nonreciprocal variations of the delay times are caused by (a) residual satellite motion, (b) by drifts of the signal delay times in the electronic components in the ground stations and in the satellites, and (c) by the difference between the uplink and downlink frequencies in combination with the propagation delay introduced by the troposphere, the ionosphere, and multipaths. All of these contributions to the delay time can be better estimated and characterised comparing the TWSTFT between remote stations to the time transfer via optical fibres between the same two remote stations.

Such optical T\&F distribution between ground stations interconnected by fibre-optic networks could furthermore provide a terrestrial backup to GNSS. In particular, the dissemination of highly accurate time signals from coastal fibre-linked stations may support resilient navigation and positioning services in low-visibility or GNSS-denied environments, including maritime applications. The work by Koelemeij et al.~\cite{Koelemeij2022} demonstrated that White Rabbit can be combined with wireless signals from a mobile-network-like infrastructure to achieve improved positioning.

The QTF-Backbone enables government labs and timing centres to test GNSS backup and PNT resilience scenarios. The needed mix of different T\&F transfer methods for resilience of the UTC network is subject to research and testing. It should be remembered that the establishment of the GNSS, which offers users a precision reference system, has opened enormous economic prospects. The European GNSS industry accounts for more than a quarter of the global GNSS market share. According to the European GNSS Agency (GSA), global revenues from GNSS devices and services are expected to grow from approximately €150 bn in 2019 to about €325 bn by 2029~\cite{GSA2019GNSSMarket}. Improved positioning systems for self-driving cars and unmanned aerial vehicles are a promising market. 

\section{Path to Implementation}
\label{sec:PathtoImplementation}

The network will initially connect high-demand science users and serve as the foundation for future expansion. The implementation of the QTF-Backbone will take several years and involves procuring fibres, equipment, local installations, and setting up the operational environment. Besides capital investment, sufficient workforce capacity is crucial. The process will be realised in phases, starting with a few PoPs to support high-impact R\&D experiments and gather insights for future phases. For technology and protocol-agnostic quantum communication, the distances between nodes should not cause more than 20 dB loss in signal strength (typically \(<\) 100~km). Such short distances are not feasible to be covered by PoPs in the short term. Therefore, these distances are bridged with ILA-S. In Table~\ref{tab:ila-spec} the requirements for choosing separations between ILA-S are summarised.

\begin{table}[h]
\caption{Requirement for a separation length between In-Line Amplifier Sites (ILA-S).}
\label{tab:ila-spec}
\small
\renewcommand{\arraystretch}{1.2}
\begin{tabular}{@{}P{0.25\textwidth} P{0.65\textwidth}@{}}
\toprule
Metric & Specification \\
\midrule
Attenuation & $\leq 21$ dB \\
Splices and connectors & Intra-span connections spliced everywhere; all connectors angled \\
Environment & Underground fibre only \\
Documentation to be supplied & (1) Bidirectional OTDR per span; (2) geographic data / cable geometry in electronic form (e.g. kmz file); (3) logical map per span identifying connectors, splices associated with changes of fibre type, and fibre type of spans or partial spans \\
\botrule
\end{tabular}
\end{table}


Preferentially, the fibre links will be routed through locations suitable to host trusted nodes, quantum repeaters and PoPs.

The choice of the QTF-Backbone´s PoPs are based on the following requirements, i.e. PoPs

\begin{itemize}

\item are at locations with the highest demand and user potential,

\item support R\&D topics defined in section~\ref{sec:RandD},

\item allow broad coverage across Germany,

\item enable connectivity to already existing fibre testbeds in federal states,

\item allow several loops, thus providing redundancy in case of fibre outages,~\cite{CLONETSDS2025Deliverables}(Deliverable 2.1, Section 2.1.4)
\item allow for their implementation and operation within the well-defined time scale of 10 years.
\end{itemize}

\begin{figure}[h]
\centering
\includegraphics[width=0.9\textwidth]{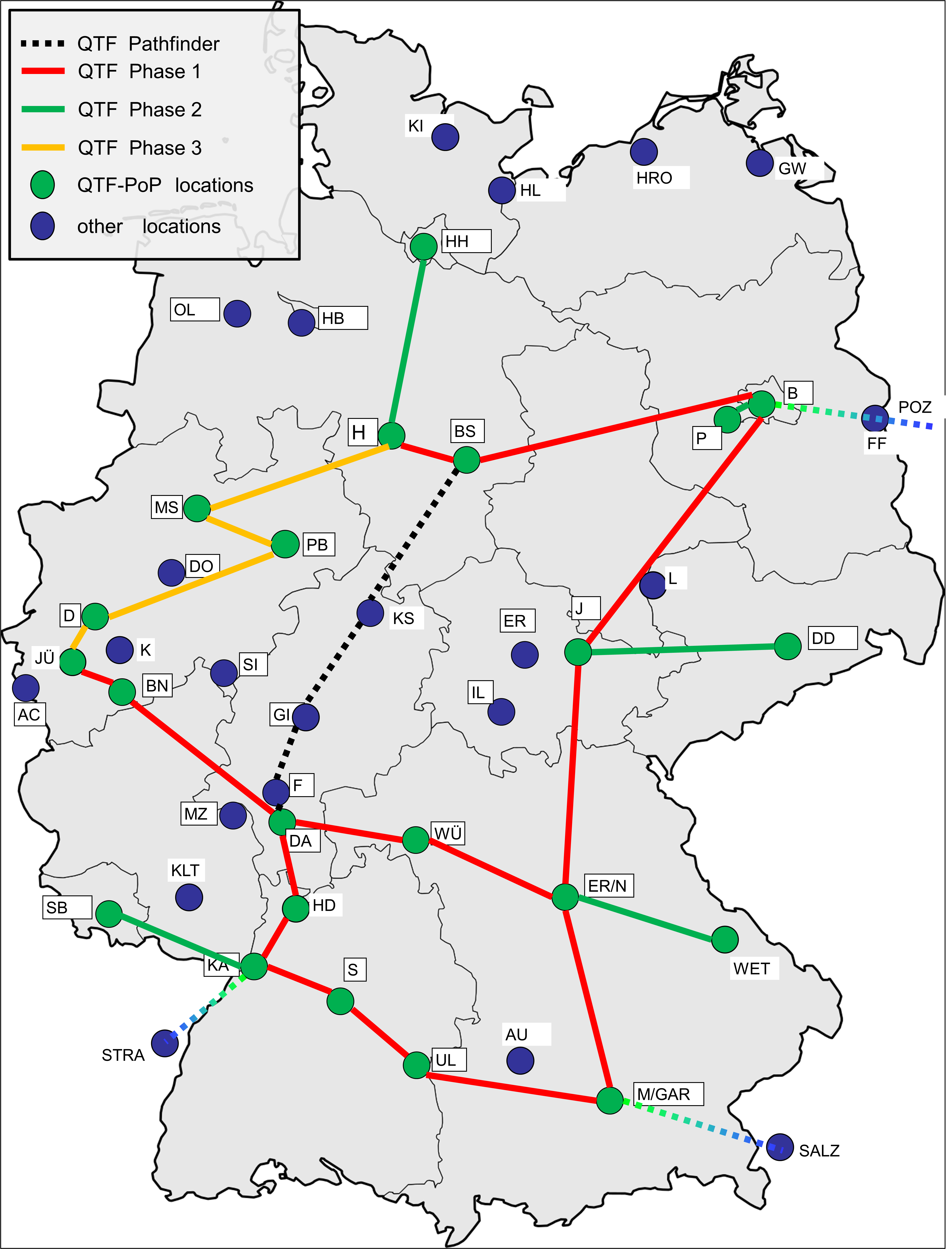}
\caption{Planning Network Phases: PoP Locations, all four proposed implementation phases are shown on this map. Link topologies and PoP locations will be refined before implementation begins. We recommend that they are updated according to the proposed R\&D projects of the stakeholders and will be decided by the RI's governance (see section~\ref{sec:Governance}).}\label{fig6}
\end{figure}

There are four phases of the installation and implementation process as shown in Figure~\ref{fig6}. In the following text and in Table~\ref{tab:pops} we summarise important parameters of the four recommended implementation phases. Due to the expected time-consuming process of fibre procurement, it is reasonable to start the procurement process as early as possible for the fibre links of proceeding phases. We stress that the four phases described here reflect the best estimate of how to maximise user reach. Link topologies and PoP locations will be refined before implementation begins. We recommend that they are updated according to the proposed R\&D projects of the stakeholders and will be decided by the RI's governance (see section~\ref{sec:Governance}).

Phase 0: This includes the so-called ``pathfinder'' link with 1 PoP (and 3 existing ILA-S). The length of the fibre links is approximately 450~km and includes part of an already existing T\&F-link from PTB to Strasburg, which plays an important role in international T\&F metrology and interlinking European NMIs in a pan-European network. It will be used for certain early tests, including additional equipment provision.

Phase 1: This first implementation phase would include 13 fully developed PoPs around Germany. The length of the installed fibre would be about 2300 km. Already in that phase certain regional quantum networks (from state initiatives (``Länderinitiativen'') or others) could be integrated, e.g. in Berlin, the Quantum Hub Thuringia fibre link from Jena to Erfurt and the application lab ``Quantum Engineering'' in Erfurt with its fibre link to Jena, or the Munich Quantum Valley (MQV) with QuKomIn from Munich to Thuringia etc., thus already in this first phase adding quantum communication stakeholders to the QTF-Backbone. Phase 1 establishes the core of the QTF-Backbone to reach the largest number of users in the shortest amount of time. It will therefore be the most technically demanding and substantial phase. We therefore anticipate that Phase 1 will be subdivided into several sub-phases over its projected time span, the details of which are beyond the scope of this work.

Phase 2: In the second implementation phase, five additional PoPs would be included; the length of the additional installed fibre would be more than 800~km.

Phase 3: The third phase would include the implementation of three further PoPs, adding approx. 700 km of installed fibre.

\begin{table}[h]
\caption{Number of PoPs and estimated link length for different implementation phases. The number of ILA-S needs to be determined in the detailed planning.}
\label{tab:pops}
\small
\renewcommand{\arraystretch}{1.2}
\begin{tabular}{@{}lll@{}}
\toprule
Phase & Number of PoPs & Fibre length (km) \\
\midrule
Pathfinder Phase-0 & 1  & 450 \\
Phase-1 & 13 & 2300 \\
Phase-2 & 5  & 850 \\
Phase-3 & 3  & 700 \\
\midrule
Sum & 22 & 4700 \\
\botrule
\end{tabular}
\end{table}


\newpage

\subsection{Funding concept}

The expansion of currently existing segments into a Germany-wide network with a European perspective will provide a unique basic infrastructure for all interested R\&D users and innovators. This requires one-off investments for the installation as well as the coverage of the rental costs of the fibre optic lines over the 10-year implementation period, until the RI can provide stable services to the users.

Existing DFN e.V. funding structures -- user and membership fees -- cannot support the QTF-Backbone's installation and pilot phases. This financial requirement can also not be met by the participating/interested research institutions during the installation and pilot phases. Federal funding will be essential during the 10-year installation and pilot period. After this initial 10-year period, when a stable level of service could be provided to users throughout Germany, it is planned that a QTF-Backbone e.V. would cover the operation costs via membership fees to the association, similar to the business model of the DFN e.V. An 'e.V.' refers to receiving the status as a registered association. Costs for the long-term fibre rental could be covered via a combination of federal and state initiatives, who oversee the QTF-Backbone via a Board of Trustees (see subsection~\ref{sec:Governance})

The below cost estimates reflect the technical design detailed in section~\ref{sec:concept} without considering potential in-kind contributions of state initiatives.

\textbf{Fibre costs:} The cost for the rental of one pair of fibres is about 650 €/km p.a., based on a DFN-internal average of a mixture in cities and overland. The real topology, i.e. the exact length of the sections within the QTF-Backbone, must be determined during the procurement process. With the topology introduced in section~\ref{sec:concept}, the rental costs for the network of two fibre pairs for the years 1 to 10 of the QTF-Backbone would be approx. 44.0 M€, see Table~\ref{tab:fibre-costs}.

\begin{table}[h]
\caption{Summary of the fibre rentals costs over the four installation and pilot phases of the QTF-Backbone RI.  *(Years 1 and 2 are covered by existing contracts.) In this estimate, the fibre optic lease for the entire QTF-Backbone was calculated, assuming a lease rate of 650 €/km per year.)}
\label{tab:fibre-costs}
\small
\renewcommand{\arraystretch}{1.2}
\begin{tabular}{@{}lllll@{}}
\toprule
Phase & Distance & Cost p.a. & Implementation duration & Implementation cost \\
\midrule
Phase 0 (pathfinder) & 450 km  & 0.6 M€ & year 3–10* & 4.7 M€ \\
Phase 1              & 2300 km & 3.0 M€ & year 2–10  & 27.0 M€ \\
Phase 2              & 850 km  & 1.1 M€ & year 4–10  & 7.7 M€ \\
Phase 3              & 700 km  & 0.9 M€ & year 6–10  & 4.6 M€ \\
\midrule
Total                &         &        &            & 44.0 M€ \\
\botrule
\end{tabular}
\end{table}


\textbf{Personnel costs:} The operation of the QTF-Backbone requires personnel with different qualifications. Central operating functions will be partly covered via existing DFN structures (e.g. DFN-NOC, DFN-AAI, etc.) within the framework of standard service agreements. For the (dynamically ongoing) implementation of the QTF-Backbone and for important overarching operational work with intensive exchange and cooperation with the end users, an additional core team will be required. The total personnel costs will be in the range of 10 full-time employees (with about 100 k€ p.a. per employee), i.e. a total of about 1 M€/year. The roles of these 10 full-time employees description is described in section~\ref{sec:Governance} and shown in figure~\ref{fig7}.

\textbf{Investment costs:} There are specific investment costs for the special devices in the QTF-Backbone in accordance with the technical design described in section~\ref{sec:concept}. These include, for example, to support quantum communication and T\&F distribution: ultra-stable lasers, masers, quartz oscillators, rf devices, transmitter \& receiver units for the provision of time signals, special bi-directional amplifiers for optical T\&F signals, stabilisation systems for the compensation of inherent runtime fluctuations, monitoring electronics, and monitoring systems.

This total investment requirement is partly subject to rapid replacement/update procurement. We include in the investment budget 1 M€ per year to be used for funding R\&D to be carried out on the QTF-Backbone as decided by the Scientific Advisory Board (SAB) and for unforeseen new technological developments.

The expenditure required by the (potentially) participating user groups from the associated R\&D institutions for their R\&D work remains completely outside the scope of the considerations here and must be provided from other funding sources via the individual R\&D projects.

All in all, the amount of funding required to operate the QTF-Backbone network infrastructure (mainly fibre costs and personnel costs) is likely to be in the medium single-digit million Euro range each year, plus the investment costs to be spread over the initial 10 years as shown in Table ~\ref{tab:capex-opex}.

\begin{table}[h]
\caption{Estimate capital expenditure (CAPEX), operational expenditure (OPEX) and total expenditure (TOTEX) for a 10-year installation and pilot phases. Estimates are based on PoPs and ILA-S and fibre links according to Table~\ref{tab:fibre-costs}.}
\label{tab:capex-opex}
\small
\renewcommand{\arraystretch}{1.2}
\begin{tabular}{@{}lllll@{}}
\toprule

\textbf{Item} & \textbf{CAPEX} & \textbf{OPEX p.a. (y1–10)} & \textbf{OPEX total} & \textbf{TOTEX} \\
\midrule
Investment & 12.6 M€ & 2.3 M€ & 22.7 M€ & 35.3 M€ \\

Personnel  & –       & 1.0 M€ & 10.0 M€ & 10.0 M€ \\

Fibre      & 44.0 M€ & –       & –       & 44.0 M€ \\

\midrule
Sum        & 56.6 M€ & 3.3 M€ & 32.7 M€ & 89.3 M€ \\
\end{tabular}
\end{table}


\textbf{Contributions of institutions:} The DFN e.V. is well suited to play a decisive role in implementing and operating the QTF-Backbone:

\begin{itemize}
    \item Plan and carry out procurement of dark fibre and equipment

\item Negotiate hosting of nodes at university IT centres

\item Plan, deploy, operate and maintain classical telecommunication links

\item Set up and run a network-operations centre taking care of monitoring the network, dealing with faults, organising access to sites etc.

\end{itemize}

PTB will feed expertise into the procurement process for T\&F equipment and will lead the deployment of T\&F services. Once operational, PTB will assist with the monitoring and maintenance of T\&F services, where specialist expertise is required. Throughout the operational lifetime of the RI, PTB will provide stable and accurate references for distribution through the network.

PTB operates a fibre link from Braunschweig to Karlsruhe, Strasbourg, and Garching for optical frequency transfer. PTB will initially contribute part of this link (Braunschweig to Frankfurt) as the ``pathfinder'' for Phase 0 to perform R\&D required to finalise the design of the service layer and install the supplementary quantum channel. It is expected that this link will ultimately be absorbed into and funded by the QTF-Backbone.

\subsection{Target groups and anticipated demand}

The primary scientific users are institutions that need infrastructure for R\&D of future quantum technologies (e.g. protocols and components) and/or stable reference frequencies for R\&D across diverse fields. Within the EU-funded projects CLONETS and CLONETS-DS, and the BMFTR-funded (previously BMBF-) QR.X project, the demands for such a network have been intensely investigated and confirmed with feedback from ca. 100 stakeholders throughout Europe and Germany, respectively. This stakeholder feedback is the basis for section~\ref{sec:RandD}.

Economic target groups will be nearly all companies that provide photonic products for the broad range of quantum technologies, either benefiting from precise reference signals, the pure access to the dark fibres as a testbed, or the development of quantum communication devices such as quantum repeaters etc. Moreover, equipment primarily needed for the installation of the backbone can be purchased from European companies (see CLONETS-DS Deliverable 2.2., Table 2.6~\cite{CLONETSDS2025Deliverables}  for more details). The QTF-Backbone provides a unique enabling technology for innovation for the quantum internet of the 21st century.

\subsection{Access models and concept for utilisation}

The QTF-Backbone aims to benefit the widest possible range of scientific and industrial users by making use of the physical signals provided by the network at specific PoPs, by using the data provided by the repository or by gaining physical access for carrying out experiments between partners in the network.

Taking this into account, the QTF Backbone will provide:

\begin{itemize}
    \item Routine operation with a high demand on continuous operation throughout the year;

\item Technical expertise indispensable for the development and operation;

\item Periods of operation spread over the year, as well as periods for maintenance and upgrades;

\item Planning of flexible and economically favourable special operations for specific scientific use cases (However, only possible to a very limited extent in terms of time after long-term planning);

\item Access to PoPs and to the data repository for co-operation partners from industrial companies and research institutes;

\item Continuity of the measurements due to traceability of T\&F references of the network to the SI-unit Second;

\item When needed, traceability to a primary frequency standard (Cs clock) or alternatively an optical clock of PTB.

\end{itemize}

It is expected that the German QTF Backbone will be used as a ‘special facility' for European metrological and quantum R\&D programmes.

User access to the QTF-Backbone will be provided with a specific quota of the usage time including a defined usage time reserved for commercial users. As outlined in the section~\ref{sec:Governance} the Scientific Advisory Board (SAB) is responsible for the selection of scientific proposals. Thus, in case of a potential oversubscription the SAB will prioritise proposals according to their scientific ranking and time demands. All information that a potential user of the backbone needs is provided on the homepage. This includes the access rules, an overview of the topology, the expected quality and availability of the backbone signals, member lists and draft co-operation agreements.

\subsection{General aspects of data utilisation}

Responsible data management is part of good research. For the generation, collection and analysis of data, timely measures need to be taken to ensure long-term storage and later reuse of the data. This means that at the start of a specific research project on the QTF-Backbone researchers must ascertain

a) which data could be relevant and

b) how these data could be stored so that they are accessible for reuse

c) to whom the data will be made accessible.

The output includes reports, scientific papers, conference presentations, training materials, deliverables. Thus, for these to be disseminated as widely as possible and used by as many stakeholders as possible, the data should be as freely accessible as possible. Each project must comply with the European Commission´s open-access data requirements.

As part of the management activities a suitable Data Management Plan (DMP) will be produced which will describe all of the data sets that will be collected, processed or generated by the project. The DMP will cover the following aspects:

\begin{itemize}
    \item the handling of research data during and after the end of the project;

   \item specification of the data that will be collected, processed or generated;

   \item the methodology and standards (including data security and ethical aspects) that will be applied;

   \item plans for data curation and preservation (including after the project).
\end{itemize}

The QTF Office will prepare a first outline of the DMP for discussion within the Executive Board and with the stakeholders to refine the DMP within a few months.  The QTF-Backbone consortium agrees to deposit its open access data sets in suitable repositories.

In order to follow current best practice on data management further information will be obtained by the consortium from the Digital Curation Centre~\cite{DCC2025}, ScienceMatters~\cite{ScienceMatters2025} and the Research Data Alliance~\cite{RDA2025}. The QTF-Backbone project will also seek to follow current best practice guidance on open data such as that from the Open Data Institute~\cite{ODI2025}.

As a minimum, the QTF-Backbone consortium will ensure that the data selected for open access:
\begin{itemize}
    \item can be linked to and is available in a standard, structured format (e.g. JSON, XML, ASCII or TIFF), so that it can be easily shared;

\item is consistently available over time, so that end users can reliably use it;

\item is stored self-descriptively or with a link to the publication/document (e.g. identified with a DOI) that accurately describes the data format and parameters used.
\end{itemize}

The selection of data to be openly accessible will be made on a case-by-case basis and agreed by the members. This will include ethical aspects and data security such as for the protection of intellectual property (IP) for any project outputs that are considered to be commercially exploitable. In such cases, it may be necessary to withhold all or some of the data generated. This will be decided by the relevant partner(s) and managed by the DMP, the Consortium Agreement and if appropriate the project's exploitation plan.

Because the QTF-Backbone utilises core communication infrastructure, some of the data associated with or generated by the infrastructure will be sensitive. For example, geolocation data of the fibre routes certainly falls into this category. Information about the operational state of the infrastructure may also be considered sensitive. Time series of fibre delay measurements enable detection and location of external disturbances, which could also be considered sensitive information. Sensitive data will be made available only to registered users on a strict need-to-know basis. Where there is a broader demand for this information, the possibility of making reduced information (e.g. with reduced spatial resolution or delayed in time) more openly available will be considered. Decisions on sensitive data will be made by the executive board after consultation with users and appropriate experts.

\subsection{Database of QTF-Backbone}

The QTF-Backbone operates its own database that stores data describing the configuration and status of the infrastructure. Some of these data are internal and serve only for the RI operation and management. At the same time, others have a broader scope and should be shared with the users.

Data that should be shared are continuously uploaded to the central database repository. This database unifies the structure and format of collected data, presents them, operates the access portal, and grants reading and writing rights to infrastructure users and the general public. The detailed specification of the database structure and list of all collected data will be designed later in the implementation phase of QTF-Backbone.

\subparagraph{Interface for Data Access}

The Repository can be accessed by two methods:
\begin{itemize}

\item A web interface is the basic method for data reading and, to a limited extent, writing. The technique is suitable mainly for static or slowly changing data. It will offer primary access to a majority of data in a human-readable form.

\item A unique Network API will be designed for automatised data reading and writing. This way, users can also build their own applications to further process and present data. This method is also suitable for a large amount of rapidly changing data.
\end{itemize}

\subparagraph{Public access}

The central web of QTF-Backbone gives the public access to the infrastructure description, list of PoPs, information for new users, contacts, etc., to serve as a primary presentation point for the general audience. Publicly available information about projects that utilise the QTF-Backbone-conducted experiments and use cases is also available. Research and the scientific community might profit from data about the achieved parameters of provided services.

\subparagraph{Restricted access}

Although the aim is to be as open as possible and allow the right to read all data that are not sensitive due to infrastructure operation or users' privacy, it is necessary to restrict access to many internal, sensitive or user-oriented data. Access to these data is granted individually and requires authentication using a password or a certificate and is dedicated to authorised users with respect to their individual privileges. Access rights will be handled by the QTF Office after consultation of the Executive Board.

\subsection{Proposed Organisational Structure and Governance}
\label{sec:Governance}

\begin{figure}[h]
\centering
\includegraphics[width=0.9\textwidth]{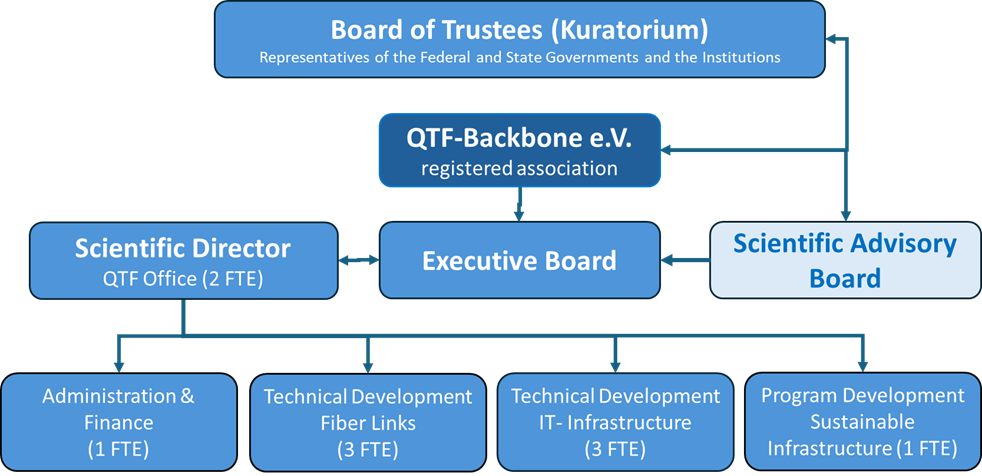}
\caption{Organisational structure of the QTF-Backbone e.V..}\label{fig7}
\end{figure}

A new non-profit association, QTF-Backbone e.V., will govern the infrastructure and oversee its nationwide deployment. We propose that the association will be supported during a ten-year installation phase by federal and state governments, after which it aims to be financially independent. Institutions contributing to the mission may join as members, with rights and obligations defined by the statutes. The Executive Board, formed by representatives of the founding institutions, is the main decision-making body, supported by a Scientific Director and the QTF Office, which handles coordination, reporting, outreach, and administration. An external SAB advises on strategy, while the Board of Trustees ensures alignment with government and funding bodies. Operations are divided across the following four departments: Administration \& Finance, Technical Development -- Fibre Links, Technical Development -- IT Infrastructure, and Program Development \& Sustainable Infrastructure. All described functions and departments are shown in Figure~\ref{fig7}.

The Executive Board is the governing body of the QTF-Backbone e.V. association. Together with the Scientific Director, who gets appointed by the Executive Board, they are responsible for steering and supervising the QTF project parts and taking decisions. The Scientific Director also acts as a link to relevant EU initiatives and is, together with the QTF Office, responsible for the overall management and coordination of the RI and outreach activities.

The QTF Office supports the Executive Board, the Scientific Director, and all members and partners of QTF-Backbone e.V. in administrative tasks and applications processes including fundraising. Together with the Scientific Director, the QTF Office is responsible for the overall management and strategic coordination of the QTF-Backbone. The QTF Office takes care of the preparation of the annual reports and the organisation of the SAB Meetings and are the first contact point for the Federal and State Governments. They oversee and coordinate the work of the operational departments described in the following:

The task of the Administration and Finance department (A\&F) is to manage the infrastructure and all financial resources. It will manage daily administration, budget oversight, and long-term strategic planning.

The Technical Development Fibre Links department (TDFL) is responsible for the design and installation of the fibre infrastructure, its operation and the feeding and securing of the signals made available in the network. It will consult participating institutes for specialised expertise as needed. Together with the IT-Development team the TDFL will operate the network operating centre (NOC).

The team supports the Department for Program Development and sustainable Infrastructure with staff training and public relations work.

The Technical Development IT Infrastructure department (TDI) is responsible for the complete IT-infrastructure of the association. This includes online monitoring of the PoP IT-infrastructure such as gateways, servers, and data recording systems, develops and maintains the data repository in line with FAIR data principles. The team will develop mapping of the complete fibre infrastructure, its current state of operation including validation flags. The team works closely with the TDFL team.

The task of the Program Development and Sustainable Infrastructure department (PD \& SI) is maintaining close contact with the knowledge institutions that are members of the association, leverages institutional strengths to advance the QTF-Backbone, foster collaboration, engage members, and coordinate outreach events. It will also support knowledge transfer in two directions: (1) to stakeholders  working with the QTF-BB-infrastructure and (2) to future stakeholders, including students who might be interested in this infrastructure for their own future work. Part of that knowledge transfer could be the establishment of a QTF-BB eAcademy, similar to the existing G{\'E}ANT eAcademy~\cite{GEANTeAcademy}. 

The team is responsible for public outreach, raising awareness, and collaboration with European organisations such as EuroQCI, or the planned ESFRI project FOREST. PD \& SI establishes and keeps contact with the European National Research and Educational Networks (NREN) and the pan-European Network GÉANT and the European Metrology Networks (EMN) as well as to the Technical Committees (TC) of the European Metrology Organisation EURAMET.

The Scientific Advisory Board advises the Board of Trustees and the Executive Board in all areas of R\&D. The advice explicitly covers the strategy and planning of R\&D work, the utilisation of results, and cooperation with national and international institutions.

The Board of Trustees provides the highest level of oversight of the QTF-Backbone to ensure that expectations of the federal ministries, state governments and institutions, which provide financial and content support to the QTF-Backbone, are represented and coordinated between all members.

\section{Conclusions and Outlook}

\subparagraph{Strengthening Germany´s position as a place for innovation in Europe and worldwide}

Germany's quantum sector has accelerated in recent years, driven by national programs~\cite{BMBF2018Quantentechnologien}, the EU's Quantum Flagship Initiative~\cite{QuantumFlagship2025}, the Horizon Europe project Qu-Test~\cite{QUTest2025} and federal and state initiatives. This tremendous support is enabling advances in photonic components, new instruments for controlling quantum systems and prototype field-deployable devices. Specifically, there are emerging innovation ecosystems for the technologies in security applications, quantum computing, quantum sensor technology, and quantum communication~\cite{DIVQSec2025} involving start-ups to big players in the semiconductor and telecommunications industries.

Furthermore, the QTF backbone will make time and frequency services that have been so far limited to national metrology institutes and a few laboratories worldwide accessible to a wider user community throughout Germany.  This will not only advance innovative research fields in Fundamental Physics, Geosciences, Astronomy and Precision metrology but also create new markets for e.g. RF measurement equipment and photonics industry.  

The QTF backbone enables Germany, at the heart of Europe, to provide R\&D users with access to the European network. Thus it will be a key infrastructure to realise and unify - literally and symbolically - Germany's and Europe's innovation goals in quantum technologies and high precision metrology. The QTF-Backbone will provide access to T\&F references at an unprecedented level of stability and accuracy, full characterisation of fibre links, and a plug-and-play infrastructure for quantum technology innovations, generating considerable synergies for all users -- from research institutions to start-ups to large companies -- and thus massively strengthening Germany as a science and technology location.

\subparagraph{Cooperation with industry and commerce}

The installation of the QTF-Backbone, with ongoing procurement of hardware, software, and methods over many years, will provide business opportunities for spin-off companies and other highly specialised innovators. User access to the QTF-Backbone with a specific quota of the usage time including usage time reserved for commercial users will become available, in the short-term, 1 to 3 years after the projects begins.  The infrastructure and services available at PoPs and via state sub-nets guarantee a low threshold even for small and medium businesses. Start-up and SMEs could access the QTF-Backbone through connections at high-tech incubators. For example, the Quantum Valley Lower Saxony High-tech Incubator (QVLS-HTI) already has a dark-fibre connection to PTB and access to frequency references and the characterisation of quantum communication components. Establishing PoPs or connections via sub-nets like this throughout Germany can be part of the third phase of implementation of the QTF-Backbone and is hence a medium-term goal.

The French REFIMEVE network illustrates how infrastructure like the QTF-Backbone can drive commercialisation. The network relies on products first developed at Université Sorbonne Paris Nord and subsequently commercialised. We anticipate that the step-by-step implementation of the QTF-Backbone would provide innovators with the opportunity to come up with superior, competing products, which could be utilised in later stages of the RI installation, with interoperability assured by early field testing using the QTF-Backbone.

\subparagraph{Contributions to national and European sovereignty, resilience and capacity to act}

Public authorities are increasingly concerned about the excessive dependence of critical infrastructures on GNSS, as GNSS radio signals and receivers are vulnerable, and backup systems are not yet in place. Since the Russian war on Ukraine, the threat to satellite-based navigation has clearly become a reality. In 2021 a UK governmental report~\cite{UKGov2017GNSS} indicated that a disruption of GNSS functionality for 24 hours would translate into an estimated loss of £1.4bn; a longer outage of 7 days was estimated at £7.6bn with applications in emergency services, maritime, and road together bearing the brunt with 87.6$\%$ of the total economic loss.

The QTF-Backbone will provide opportunities to test fibre-based synchronisation technologies as an alternative to GNSS at a much larger scale than previously possible. Critical infrastructures, such as the Galileo control centre Oberpfaffenhofen, can be connected and will then be able to gain experience in evaluating these technologies.

Fibre-based synchronisation benefits both timing and future terrestrial navigation systems. In a recent Science for Policy report~\cite{JRC2023QuantumComms}, the Joint Research Centre of the European Commission has assessed alternative positioning, navigation, and timing technologies for potential deployment in the EU. Notably, fibre-based time transfer featured in over half the technologies evaluated.

\section{List of abbreviations}

A\&F		Administration and Finance Department

API		Application Programming Interface

AQUnet	Austrian Quantum Fibre Network

ASCII		American Standard Code for Information Interchange

BDBOS	Federal Agency for Public Safety Digital Radio

BKG		Federal Agency for Cartography and Geodesy

BMBF		Federal Ministry for Education and Research

BMFTR		Federal Ministry of Research, Technology and Space

BMI		Federal Ministry of the Interior and Community

BOOSTED	Belgium Optical network for Optical frequency Standards and TimE Dissemination

BQC		Bline Quantum Computation

CITAF		Czech Infrastructure for Time and Frequency

CLONETS	CLOck NETwork Services Project

CLONETS-DS	Design Study for a Europe-wide Metrology Network

COTS		Commercial Off-The-Shelf

CPT		Charge, Parity and Time

C-TFN		GÉANT Core Time/Frequency Network

CV-QKD	Continuous Variable Quantum Key Distribution

DAS		Distributed Acoustic Sensing

DE-CIX	German Commercial Internet Exchange

DFN		Association of German Research Network (DFN e.V.)

DI		Device Independent

DMP		Data Management Plan

DOI		Digital Object Identifier

DORIS		Doppler Orbitography and Radiopositioning Integrated by Satellite

DV-QKD	Discrete Variable Quantum Key Distribution

DWDM		Dense Wavelength Division Multiplexing

EC		European Commission

ESFRI		European Strategy Forum on Research Infrastructures

EuroQCI	European Quantum Communication Infrastructure

e.V.		registered association (in German: eingetragener Verein)

FAIR		Findable, Accessible, Interoperable and Reusable

FOREST	Fibre-based Optical netwoRk for European Science and Technology

FNQ		Fraunhofer Competence Network for Quantum Computing

GÉANT		Collaboration of European National Research and Education Networks

GFZ		German Research Centre for Geosciences

GLONASS	Global'naya Navigatsionnaya Sputnikovaya Sistema

GNSS		Global Navigation Satellite Systems

GPS		Global Positioning System

GRACE-FO	Gravity Recovery and Climate Experiment Follow-On

GSA		GNSS Agency

HTAD    Hightech-Agenda-Deutschland 

IAG		International Association of Geodesy

ICRF		International Celestial Reference Frame

IHRF	   	International Height Reference Frame

ILA-S		In-line Amplifier Site

IPPP		Integer Precise Point Positioning

IQB		Italian Quantum Backbone

IRU		Indefeasible Right of Use

ITRF		International Terrestrial Reference System and Frame

JSON		Javascript Object Notation

KTN		Core Transport Network of the Federal Government in Germany

LIFT		Italian Link for Frequency and Time

MDI-QKD	Measurement Device Independent Quantum Key Distribution

MHz		Megahertz

MPQ		Max-Planck-Institute for Quantum Optics

MQV		Munich Quantum Valley

NDFF		National Dark Fibre Facility (UK)

NLPQT		National Laboratory for Photonics and Quantum Technologies (Poland)

NMI		National Metrology Institute

NOC		Network Operating Centre

NPL		National Physical Laboratory (UK)

NREN		National Research and Education Network

NTC		National Timing Centre (UK)

OCXO    Oven Controlled Crystal Oscillators

OFDR		Optical Frequency Domain Reflectometry

OTDR		Optical Time-Domain Reflectometer

PD\&SI		Program Development and Sustainable Infrastructure

PNT		Positioning, Navigation, and Timing

PoP		Point of Presence

PPP-AR	Precise Point Positioning with Ambiguity Resolution

PPS		Pulse Per Second

PSNC		Pozna\'{n} Supercomputing and Networking Centre

PTB		Physikalisch-Technische Bundesanstalt

QKD		Quantum Key Distribution

QTF		Quantum, Time and Frequency

QuKomIn	Quantum Communication Infrastructure Bavaria

QVLS-HTI	Quantum Valley Lower Saxony High-Tech Incubator

R\&D		Research and Development

R\&E		Research and Education

REFIMEVE	Réseau Fibré Métrologique à Vocation Européenne Network

RI		Research Infrastructure

SAB		Scientific Advisory Board

SaSER		Save and Secure European Routing Test-net

SI		Système International d'Unités

SLR		Satellite Laser Ranging

SME		Small and medium enterprises

TC		Technical Committees

TDFL		Technical Development Fibre Links

TDI		Technical Development IT Infrastructure

T\&F		Time and Frequency

THz		Terahertz

TIFF		Tagged Image File Format

TiFOON	Time and Frequency over Optical Networks Project

TRL		Technology Readiness Level

TRUST   Transparency, Responsibility, User focus, Sustainability, and Technology

TWCP		Two-Way Carrier-Phase satellite time transfer

TWSTFT	Two-Way Satellite Time and Frequency Transfer

UniBW		University of the Bundeswehr

UTC		Universal Time Coordinated

VLBI		Very Long Baseline Interferometry

VGOS		VLBI Global Observing System

WDM		Wavelength Division Multiplexing

XML		Extensible Markup Language

X-WiN		German Science Network

\subsection{Acknowledgements}

We thank the community for their tremendous support for the QTF-Backbone initiative in the form of feedback and corrections to this manuscript, LoIs, and discussions. We thank the partners throughout Europe who have paved the way with optical fibre backbones for quantum communication and time and frequency distribution. We gratefully acknowledge valuable discussions, feedback, and support from our broader community. 
The author TCL is the corresponding author for this paper and was responsible for the final editing of the manuscript. The following authors are the core contributors to the proposal of the QTF-Backbone including its technical design, integration, impact, concepts for funding and utilisation, and governance: KB, PK, JK, SK, TCL, DM, SN-J, SSch and HS. These authors served as experts and played an important role in the drafting of the Transformative R\&D Topics: CB, JB, UB, DBud, CD, THei, SH, RHol, UH, NH, CI, TKlu, MKra, J, DP, LSan, SSte. All authors read and approved the final manuscript.

This version of the article has been accepted for publication, after peer review (when applicable) but is not the Version of Record and does not reflect post-acceptance improvements, or any corrections. The Version of Record is
available online at: https://doi.org/10.1140/epjs/s11734-026-02451-3

\bibliographystyle{sn-mathphys-num}
\bibliography{sn-bibliography_fixed}

\end{document}